\documentclass[aps,showpacs,showkeys,amsmath,amssymb,twocolumn,
floatfix,superscriptaddress]{revtex4}
\usepackage{graphicx}
\usepackage{amssymb}
\usepackage{dcolumn}
\usepackage{bm}
\usepackage{color}
\begin{document}

\title{Formation of bremsstrahlung in an absorptive QED/QCD medium} 

\author{M.~Bluhm}
\email{mbluhm@ncsu.edu}
\affiliation{SUBATECH, UMR 6457, Universit\'{e} de Nantes, 
Ecole des Mines de Nantes, IN2P3/CNRS. 4 rue Alfred Kastler, 
44307 Nantes cedex 3, France}
\author{P.~B.~Gossiaux}
\affiliation{SUBATECH, UMR 6457, Universit\'{e} de Nantes, 
Ecole des Mines de Nantes, IN2P3/CNRS. 4 rue Alfred Kastler, 
44307 Nantes cedex 3, France}
\author{T.~Gousset}
\affiliation{SUBATECH, UMR 6457, Universit\'{e} de Nantes, 
Ecole des Mines de Nantes, IN2P3/CNRS. 4 rue Alfred Kastler, 
44307 Nantes cedex 3, France}
\author{J.~Aichelin}
\affiliation{SUBATECH, UMR 6457, Universit\'{e} de Nantes, 
Ecole des Mines de Nantes, IN2P3/CNRS. 4 rue Alfred Kastler, 
44307 Nantes cedex 3, France}


\keywords{radiative energy loss, radiation formation time, damping of radiation, 
density effect, LPM effect, Ter-Mikaelian effect}
\pacs{12.38.Mh, 25.75.-q, 52.20.Hv} 

\begin{abstract}
The radiative energy loss of a relativistic charge in a dense, 
absorptive medium can be affected significantly by damping 
phenomena. The effect is more pronounced for large energies of 
the charge and/or large damping of the radiation. This can be 
understood in terms of a competition between the formation time 
of bremsstrahlung and a damping time scale. We discuss this 
competition in detail for the absorptive QED and QCD medium, 
focusing on the case in which the mass of 
the charge is large compared to the in-medium mass of the 
radiation quanta. We identify the regions in energy and parameter 
space, in which either coherence or damping effects are of major 
importance for the radiative energy loss spectrum. We show that 
damping phenomena can lead to a stronger suppression of the 
spectrum than coherence effects.
\end{abstract}

\maketitle

\section{Introduction \label{sec:1}}

The quenching of jets associated with the strong suppression of the yields of 
high transverse momentum hadrons in comparison with proton-proton collisions 
has been observed experimentally in relativistic heavy-ion collisions at the 
Relativistic Heavy Ion Collider at BNL~\cite{Adcox02,Adler02} 
and at the Large Hadron Collider at CERN, cf.~\cite{Muller} for a recent review. 
Originally, such a signal was predicted by Bjorken~\cite{Bjorken} as an evidence 
for the formation of a deconfined plasma state of QCD matter, in which 
propagating partons suffer from an enhanced in-medium energy loss. 

Considered as dominant contribution, the radiative energy loss of relativistic 
partons due to medium-induced gluon radiation was, for instance, studied 
in~\cite{Gyulassy94,Wang95,Baier95,Baier97,Zakharov,GLV,AMY,ASW,Dokshitzer}. 
Important for the discussion of bremsstrahlung in a dense medium are possible 
coherence effects. These lead to a suppression of the radiation spectrum 
compared to the spectrum due to incoherent scatterings. This was first realized 
by Landau, Pomeranchuk~\cite{Landau53} and Migdal~\cite{Migdal56} who studied in 
QED the possibility of a destructive interference of radiation amplitudes, 
which stem from multiple scatterings of an electric charge traversing dense 
matter, within the formation time of radiation (LPM effect). This effect was 
later generalized to QCD in~\cite{Baier95,Baier97}. 

Likewise important are modifications of the radiation spectrum due to the 
dielectric polarization of the medium, which is known as the Ter-Mikaelian 
(TM) effect~\cite{Ter-Mikaelian}. The medium polarization results in a change of 
the dispersion relation of the radiated quanta. In this way, the spectrum becomes 
regulated in the soft and collinear regions. In QCD, the TM effect on the 
radiation spectrum was investigated by considering bremsstrahlung gluons with 
a finite in-medium mass~\cite{Kampfer00,Djordjevic03}. 

The concept of a formation time (or length) of radiation turned out to be extremely 
fruitful for the discussion of radiative energy losses~\cite{Baier95}. It allows 
for a semi-quantitative understanding of the pattern of the radiation 
spectra. The spectrum reduction due to the LPM effect, for example, was 
qualitatively analyzed in this way for QED in~\cite{Galitsky,Feinberg} and for QCD 
in the review~\cite{BSZ}: If the radiation formation length, which depends on 
the energy $\omega$ of the emitted quantum and on the properties of the 
dense medium (e.g. the transport coefficient $\hat{q}$), 
is large compared to the mean free path in the medium, multiple 
scatterings will contribute coherently to the emission of a single radiation quantum. 
This represents the LPM regime of coherent radiation. 
In the opposite case, radiation quanta will be formed independently at each individual 
scattering. This represents the Bethe-Heitler (BH) regime of incoherent radiation. 

In this work, we extend this analysis in order to include damping phenomena 
within an absorptive medium into the discussion. The latter are responsible for 
a reduction of the radiative energy loss of the charge and the associated radiative energy loss spectrum 
as was recently studied in~\cite{Bluhm11}. We stress that this behaviour is not 
a simple consequence of the rather trivial reduction of any radiation spectrum in the far distance due 
to the absorption of already formed radiation but a result of the effect which damping 
has on the creation of the radiation itself. 
Just as the TM effect alters the radiation probability in 
the soft part of the spectrum~\cite{BaierKatkov}, damping mechanisms lead 
also to a modification of the probability for emitting bremsstrahlung. 
This can be understood by viewing the formation of bremsstrahlung to be hampered by 
damping effects, in particular, when formation times become large. 

The article is organized as follows: In Sec.~\ref{sec:2}, our results~\cite{Bluhm11} 
for the radiative energy loss spectrum per unit length of a relativistic charge 
in an absorptive, electro-magnetic plasma are reviewed. We highlight that for a 
semi-quantitative discussion of this spectrum both the radiation formation time, 
as discussed in Sec.~\ref{sec:3_1}, and the 
time scale associated with damping effects are important. Their competition is 
analyzed in detail in Sec.~\ref{sec:3_2}. This allows for an identification of the 
regions in energy and parameter space, in which damping phenomena are of importance. 
A physical discussion in terms of the spectra is given in Sec.~\ref{sec:3_3}. 
In Sec.~\ref{sec:4}, we apply the methods of Sec.~\ref{sec:3} in order to study phenomenologically the influence 
of damping effects on the gluon bremsstrahlung spectrum in the hot QCD plasma and 
advocate some possible physical consequences in the 
conclusions in Sec.~\ref{sec:5}, where our results are summarized. 
Throughout this work natural units are 
used, i.~e.~$\hbar=c=1$ with $\hbar c\simeq 0.197$ GeV$\cdot$fm. 

\section{Radiative energy loss spectrum in an absorptive plasma \label{sec:2}}

The radiative energy loss spectrum per unit length of an asymptotic, relativistic charge 
$q$ with energy $E$ and mass $M$ (Lorentz-factor $\gamma=E/M$) traversing a 
polarizable and absorptive plasma was found in~\cite{Bluhm11} to be 
substantially reduced by both medium polarization and damping effects. This study was performed for an electro-magnetic plasma, where small fractional photon energies $\omega/E$ were considered. In~\cite{Bluhm11}, 
the dense medium was modelled by a 
complex squared index of refraction, 
\begin{equation}
n^2(\omega)=1-\frac{m^2}{\omega^2}+\frac{2i\Gamma}{\omega},
\end{equation} 
with $m$ and $\Gamma$ accounting for the in-medium 
mass and damping rate of the radiation quanta, respectively. 

Considering an infinite medium with permeability $\mu(\omega)=1$, one finds for the radiative energy 
loss spectrum for positive $\omega$ 
\begin{eqnarray}
\nonumber
 -\frac{d^2W}{dzd\omega} & \simeq & \frac{\alpha}{6\pi} \frac{\hat{q}\omega}{E^2} 
 \int_0^\infty d\bar{t}\,\mathcal{F}(\bar{t}\,) \\
\label{equ:1}
 & & \hspace{3mm} 
 \times \sin\left[\omega\bar{t}\left(1-\vert n_r\vert\beta\right)+
 \frac{\omega\vert n_r\vert\beta\hat{q}}{12E^2}\bar{t}^{\,2}\right] 
 \end{eqnarray}
in linear response theory~\cite{Bluhm11}. Here, $\alpha=q^2/(4\pi)$ is the coupling, 
$\beta=\sqrt{1-1/\gamma^2}\simeq 1-1/(2\gamma^2)$, 
$n_r$ is the real part of 
$n(\omega)=n_r(\omega)+i n_i(\omega)$ and $\hat{q}$ denotes the mean 
accumulated transverse momentum squared of the deflected charge per unit 
time. We note that in~\cite{Bluhm11} $\hat{q}$ denoted only one-half of the mean accumulated transverse momentum squared of the deflected charge per unit time, see also~\cite{Bluhm12}. 

The factor $\mathcal{F}(\bar{t}\,)$ in Eq.~(\ref{equ:1}) enters the spectrum for every charge trajectory defining $\Delta\vec{r}\equiv\vec{r}(t)-\vec{r}(t')$ with $\bar{t}=t-t'$ and reads $\mathcal{F}\equiv \exp[-\omega|n_i|\Delta r]$, cf.~\cite{Bluhm11,Bluhm12}. Since $\Delta r\geq 0$, it introduces a genuine damping factor in the spectrum Eq.~(\ref{equ:1}) for an absorptive medium which is related to the absolute value of the imaginary part of $n(\omega)$. A specific form for $\mathcal{F}(\bar{t}\,)\simeq\exp[-\omega|n_i|\beta \bar{t}\,(1-\hat{q} \bar{t}/(12E^2))]$ was used in~\cite{Bluhm11}. This form is a consequence of the particular trajectory considered in~\cite{Landau53} yielding $\langle\Delta r\rangle\simeq\beta\bar{t}\, [1-\hat{q}\bar{t}/(12E^2)]$, which takes explicitly into account the effect of multiple scatterings in the approximation of small deflection angles accumulated within the time-duration $\bar{t}$. 
The made approximation is, however, strictly valid only for $\bar{t}\ll t_\mathrm{diff}=6E^2/\hat{q}$~\cite{Bluhm12}. For larger times a different dependence than the one used in~\cite{Bluhm11} is to be expected since, physically, $\mathcal{F}(\bar{t}\,)$ should decrease monotonically. Since for $\bar{t}\ll t_\mathrm{diff}$ the term quadratic in $\bar{t}$, being proportional to $\bar{t}/t_\mathrm{diff}$, represents only a minor correction to $\omega|n_i|\beta\bar{t}$ in $\mathcal{F}(\bar{t}\,)$ and since $t_\mathrm{diff}$ is indeed a very large time scale for relativistic charges, in the following it suffices to consider only the linear time-dependence in the exponential damping factor, i.e. 
\begin{equation}
\label{equ:DampingFactor}
 \mathcal{F}(\bar{t}\,)\simeq\exp[-\bar{t}/t_d] \,,
\end{equation}
where we define 
\begin{equation}
\label{equ:DampingTime}
 t_d\simeq(\omega\vert n_i\vert\beta)^{-1}
\end{equation}
as damping time scale.

In the limit $n_r=1$ and $n_i=0$, Eq.~(\ref{equ:1}) becomes 
\begin{equation}
\label{equ:2}
 -\frac{d^2W}{dzd\omega} \simeq \frac{\alpha}{6\pi} \frac{\hat{q}\omega}{E^2} 
 \int_0^\infty d\bar{t}\,\sin\left[\omega\bar{t}\left(1-\beta\right)+
 \frac{\omega\beta\hat{q}}{12E^2}\bar{t}^{\,2}\right] ,
\end{equation}
which agrees with the result for the radiation spectrum reported 
in~\cite{Landau53} if $\hat{q}$ is properly identified with the parameters 
used therein. Substituting in Eq.~(\ref{equ:2}) $\bar{t}$ by $u/[\omega(1-\beta)]$, one obtains 
\begin{equation}
\label{equ:2'}
 -\frac{d^2W}{dzd\omega} \simeq \frac{\alpha\hat{q}}{3\pi M^2} \int_0^\infty du \sin\left[u+\mathcal{A}u^2\right] \,,
\end{equation}
where $\mathcal{A}=\hat{q}E^2/(3\omega M^4)$. The latter integral may be evaluated analytically in terms of the known Fresnel-integrals $\mathcal{S}(y)$ and $\mathcal{C}(y)$ resulting in 
\begin{eqnarray}
\nonumber
 -\frac{d^2W}{dzd\omega} & \simeq & \frac{\alpha\hat{q}}{6\pi M^2} \sqrt{\frac{\pi}{2\mathcal{A}}} \Bigg\{ \cos\left(\frac{1}{4\mathcal{A}}\right)\left[1-2\mathcal{S}(y)\right] \\
\label{equ:2''}
 & & + \sin\left(\frac{1}{4\mathcal{A}}\right)\left[2\mathcal{C}(y)-1\right]\Bigg\}
\end{eqnarray}
with $y^{-1}=\sqrt{2\pi\mathcal{A}}$. For small $\mathcal{A}$, one finds formally from Eq.~(\ref{equ:2''}) that $d^2W\simeq d^2W_{BH}$ with
\begin{equation}
\label{equ:3}
 -\frac{d^2W_{BH}}{dzd\omega} = \frac{\alpha\hat{q}}{3\pi M^2} \,, 
\end{equation}
cf.~\cite{BluhmQM11}, which is equivalent to the BH-result for the bremsstrahlung spectrum 
from incoherent scatterings, cf.~e.g.~\cite{Galitsky,Feinberg,Peigne}. More precisely, the relative deviation of Eq.~(\ref{equ:3}) from Eq.~(\ref{equ:2'}) is less than $10\%$ for $\mathcal{A}<0.026$. 
This case is achieved for large $\omega$, whenever $E\ll 3M^4/\hat{q}$. The $\omega$-independent result in Eq.~(\ref{equ:3}) provides a particularly suitable reference point for an analysis of in-medium effects on the radiative energy loss spectrum. 

In~\cite{Bluhm11}, it has been shown that for non-zero $n_i$ the differential spectrum in Eq.~(\ref{equ:1}) is reduced with increasing $E$ (for fixed $\Gamma$) or with increasing $\Gamma$ (for fixed $E$).
This is a consequence of the 
increasing influence of $\mathcal{F}(\bar{t}\,)$ in Eq.~(\ref{equ:1}) with increasing 
$\Gamma$, but also with decreasing $\omega/E$, as was discussed in~\cite{BluhmQM11}. The behaviour of the spectrum can be understood by analyzing the competition of two different time scales: The formation time of radiation for negligible damping $t_f$, and the damping time, $t_d$, at which the radiation 
amplitude in Eq.~(\ref{equ:1}) is essentially suppressed due to $\mathcal{F}(\bar{t}\,)$. 

The formation time of radiation in the limit $\Gamma\ll m\ll\omega$, i.e.~in the case of negligible damping effects, can be found, following~\cite{Landau53}, from a condition for the phase of the oscillating function in Eq.~(\ref{equ:1}) reading 
\begin{equation}
\label{equ:X}
 t_f \left[\omega-k(\omega)\beta\right] + t_f^2 \frac{\hat{q}k(\omega)\beta}{12E^2} 
 \simeq 1 
\end{equation}
with $k(\omega)=\sqrt{\omega^2-m^2}$. The first term, which is linear in $t_f$, is specific for a single scattering process, while the second term, which is quadratic in $t_f$, is genuine for multiple scatterings~\cite{Landau53}.
Considering in Eq.~(\ref{equ:X}) either the linear or the quadratic term in $t_f$ as the dominant contribution, $t_f$ may be estimated by $t_f\simeq\min\{t_f^{(s)}, t_f^{(m)}\}$, i.e.~the minimum of two time scales $t_f^{(s)}$ (the photon formation time in incoherent (single) scatterings) and $t_f^{(m)}$ (the photon formation time in coherent (multiple) scatterings). These time scales are defined as 
\begin{eqnarray}
\label{equ:tfsEDomega}
 t_f^{(s)} & \simeq & \frac{1}{\omega-k(\omega)\beta} \,, \\
\label{equ:tfmEDomega}
 t_f^{(m)} & \simeq & \sqrt{\frac{12E^2}{\hat{q} k(\omega)\beta}} \,.
\end{eqnarray}

With these time scales, it is possible to rewrite Eq.~(\ref{equ:1}) normalized with respect to the BH-result in Eq.~(\ref{equ:3}) as 
\begin{eqnarray}
\nonumber
 & & \hspace{-8mm} -\frac{d^2W}{dz d\omega} \left/ \left(-\frac{d^2W_{BH}}{dz d\omega}\right)\right. \simeq \\
\label{eq:spectrum_ratio}
 & & \hspace{-2mm} \frac{1}{t_{BH}} 
 \mathrm{Im} \int_0^{\infty} d\bar{t}\, \exp\left[-\frac{\bar{t}}{t_d}+i\left(\frac{\bar{t}}{t_f^{(s)}}+
 \frac{\bar{t}^{\,2}}{t_f^{(m)\,2}}\right)\right]
\end{eqnarray}
with $t_{BH}\simeq 2\gamma^2/\omega$. The integral in Eq.~(\ref{eq:spectrum_ratio}) can be evaluated exactly~\cite{Gradshteyn}, leading to 
\begin{eqnarray}
\nonumber
 & & \hspace{-8mm} -\frac{d^2W}{dz d\omega}\left/ \left(-\frac{d^2W_{BH}}{dz d\omega}\right)\right. \simeq \\ 
\label{eq:exact_spectrum}
 & & \hspace{-2mm} \frac{t_f^{(m)}}{2t_{BH}} \mathrm{Im}
 \left[\sqrt{i\pi}e^{i\zeta^2}\left(1-\mathrm{erf}(\sqrt{i}\zeta)\right)\right]\,,
\end{eqnarray}
with $\zeta=(1/t_d-i/t_f^{(s)})\,t_f^{(m)}/2$ and erf$(\xi)=\sqrt{4/\pi}\int_0^\xi du\,e^{-u^2}$. This result is valid for either $1/t_d\neq 0$ or $1/t_f^{(m)}\neq 0$, while for $1/t_d=1/t_f^{(m)}=0$ the integral in Eq.~(\ref{eq:spectrum_ratio}) needs additional regulation. 

The behaviour of the full spectrum relative to the BH-result can be reproduced semi-quantitatively by examining the limiting expressions of Eq.~(\ref{eq:exact_spectrum}) obtained if one of the time scales $t_d$, $t_f^{(s)}$ or $t_f^{(m)}$ is much smaller than the other two. In the case $t_d\gg t_f^{(s)}$ and $t_d\gg t_f^{(m)}$, which comprises the special case $t_d\to\infty$, one finds from Eq.~(\ref{eq:exact_spectrum}) 
\begin{equation}
\label{eq:small_damping}
 -\frac{d^2W}{dz d\omega}\left/ \left(-\frac{d^2W_{BH}}{dz d\omega}\right)\right. \simeq\kappa\frac{t_f}{t_{BH}}
\end{equation}
with $\kappa=1$ if $t_f^{(s)}\ll t_f^{(m)}$ and $\kappa=\sqrt{\pi/8}$ if $t_f^{(m)}\ll t_f^{(s)}$ instead. This particular limit comprises the LPM effect as discussed in~\cite{Galitsky}. Accordingly, we define 
\begin{equation}
-d^2W_{LPM}\simeq -d^2W_{BH}\,\sqrt{\frac{\pi}{8}}\,\frac{t_f^{(m)}}{t_{BH}} \,.
\label{eq:def_d2WLPM}
\end{equation} 
Moreover, it turns out that Eq.~(\ref{eq:small_damping}) is also valid if $t_d$ is large compared to either $t_f^{(s)}$ or $t_f^{(m)}$ only. 

In case $t_d$ is small with respect to $t_f^{(s)}$ and $t_f^{(m)}$, one may expand the integrand in Eq.~(\ref{eq:spectrum_ratio}) formally as 
\begin{eqnarray}
\nonumber
 & & \hspace{-8mm} \mathrm{Im}\, \left\{\exp\left[-\frac{\bar{t}}{t_d}+i\left(\frac{\bar{t}}{t_f^{(s)}}+
 \frac{\bar{t}^{\,2}}{t_f^{(m)\,2}}\right)\right]\right\} = \\
 & & e^{-\bar{t}/t_d}\left\{\frac{\bar{t}}{t_f^{(s)}}+\frac{\bar{t}^{\,2}}{t_f^{(m)\,2}}+\mathcal{O}(\bar{t}^{\,3})\right\}
\end{eqnarray}
and evaluate the simple integrals realizing that higher-order terms in $\bar{t}$ give rise to subdominant contributions. In the regime, in which $t_d\ll t_f^{(s)}\ll t_f^{(m)}$, one finds 
\begin{equation}
\label{eq:big_damping_small_ts}
 -\frac{d^2W}{dz d\omega} \left/ \left(-\frac{d^2W_{BH}}{dz d\omega}\right)\right. \simeq \frac{t_d^2}{t_f^{(s)}\,t_{BH}} \,.
\end{equation}
In case $t_d\ll t_f^{(m)}\ll t_f^{(s)}$, one has to distinguish between $t_d\ll t_f^{(m)\,2}/t_f^{(s)}$, in which case Eq.~(\ref{eq:big_damping_small_ts}) is also found, and $t_d\gg t_f^{(m)\,2}/t_f^{(s)}$, which leads to 
\begin{equation}
\label{eq:big_damping_large_ts}
 -\frac{d^2W}{dz d\omega} \left/ \left(-\frac{d^2W_{BH}}{dz d\omega}\right)\right. \simeq \frac{2\,t_d^3}{t_f^{(m)\,2}\,t_{BH}} \,.
\end{equation}

\begin{figure}[t]
\centering
\includegraphics[width=0.45\textwidth]{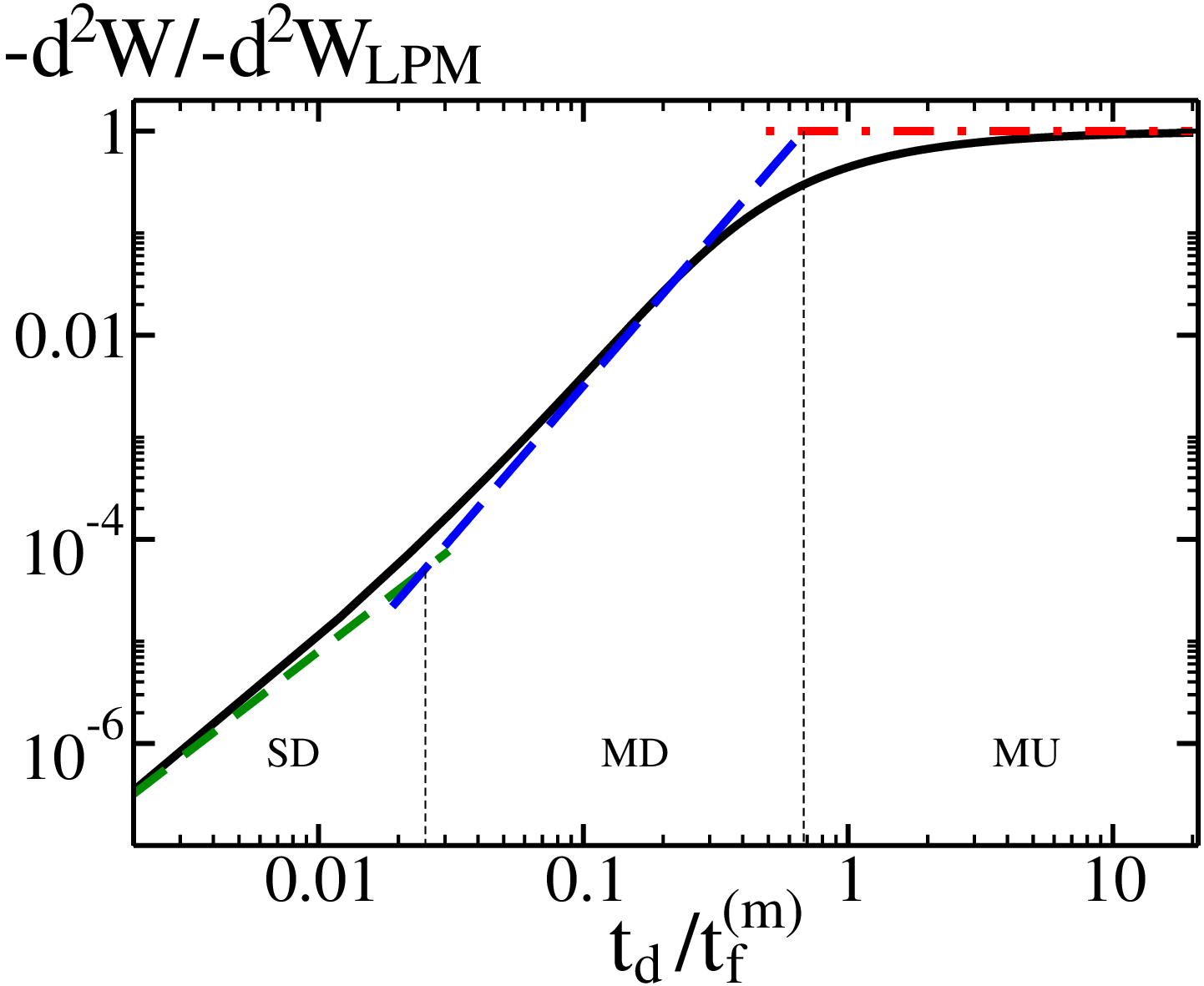}
\caption[]{\label{Fig:0new} (Color online) The ratio $(-d^2W)/(-d^2W_{LPM})$ -- that is Eq.~(\ref{equ:1}) divided by  Eq.~(\ref{eq:def_d2WLPM}) -- as a function of $t_d/t_f^{(m)}$ (solid curve) illustrating the impact of damping effects on coherence effects in an absorptive medium. Here, the special case of $t_f^{(m)}$ small compared to $t_f^{(s)}$ is considered, 
taking $t_f^{(s)}\equiv t_{\rm BH}=20\,t_f^{(m)}$, see text for details. For large $t_d/t_f^{(m)}$, damping mechanisms do not significantly affect the spectrum. For comparison, the limiting expressions related to Eqs.~(\ref{eq:big_damping_small_ts}),~(\ref{eq:big_damping_large_ts}) and~(\ref{eq:small_damping}) (short-dashed curve at small $t_d/t_f^{(m)}$, long-dashed curve at intermediate $t_d/t_f^{(m)}$ and dash-dotted curve at large $t_d/t_f^{(m)}$, respectively) are highlighted as well. The vertical lines separate regions characterized by different hierarchies in time scales, where the acronyms SD, MD and MU are explained below in Fig.~\ref{Fig:2newPLUS} and Sec.~\ref{sec:3_3}. }
\end{figure}

In order to highlight the influence of an absorptive medium on coherence effects in that medium, one may normalize the full spectrum in Eq.~(\ref{equ:1}) rather with respect to the spectrum expression $(-d^2W_{LPM})$ relevant in the regime, in which coherence effects dominate ($t_f^{(m)}\ll t_f^{(s)}$). 
The quantity $(-d^2W)/(-d^2W_{LPM})$ is shown in Fig.~\ref{Fig:0new} as a function of $t_d/t_f^{(m)}$ for the case $t_f^{(m)}=t_f^{(s)}/20$. As evident from Fig.~\ref{Fig:0new}, the full spectrum is significantly suppressed compared to $(-d^2W_{LPM})$ for small $t_d$, i.e. when damping effects are large. Moreover, Fig.~\ref{Fig:0new} demonstrates that the limiting expressions in Eqs.~(\ref{eq:small_damping}),~(\ref{eq:big_damping_small_ts}) and~(\ref{eq:big_damping_large_ts}) account fairly well for the behaviour of the radiative energy loss spectrum in an absorptive medium in the different physical regimes.

The above scaling laws show impressively the importance of the parameter-sensitive time scales $t_f^{(s)}$, $t_f^{(m)}$ and $t_d$: From their interplay, the structure of the radiative energy loss spectrum in comparison with the BH-result as reference spectrum can be obtained semi-quantitatively. In the next section, we will study these time scales and their competition in detail and discuss the meaning of our findings for the spectra. 

\section{Analysis of the regimes in the QED-case \label{sec:3}}

\subsection{Time scales in the absence of damping \label{sec:3_1}}

\begin{figure}[t]
\centering
\vspace{1.5mm}
\includegraphics[width=0.45\textwidth]{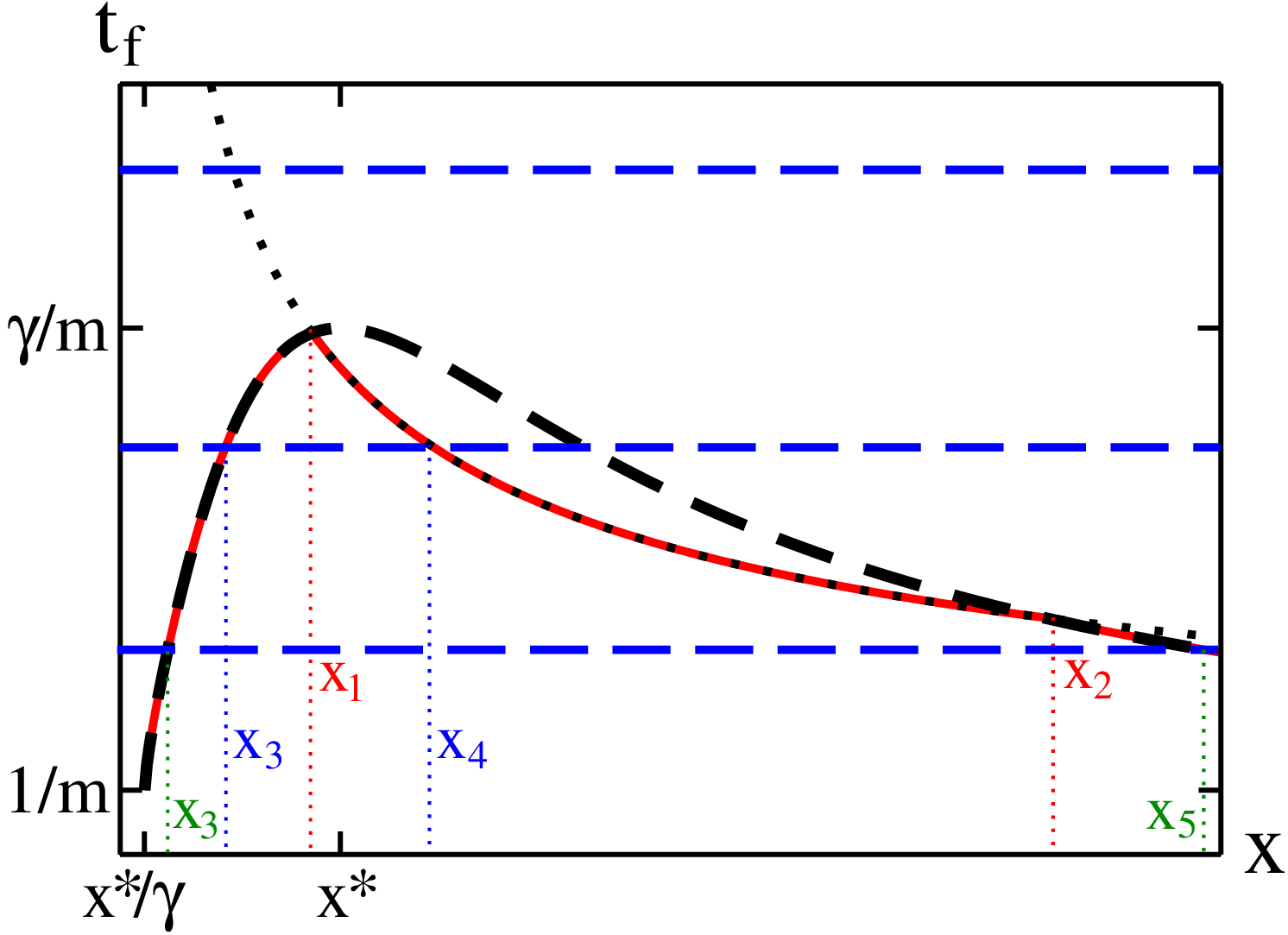} 
\caption[]{\label{Fig:1new} (Color online) Visualization of the radiation formation time $t_f$ 
as a function of $x$ for selected parameter values and fixed $\gamma$. 
Short-dashed and dotted curves exhibit $t_f^{(s)}$ and $t_f^{(m)}$, 
respectively, as defined in Eqs.~(\ref{equ:tfsED}) and~(\ref{equ:tfmED}). 
The solid curve depicts our estimate for $t_f$ given by the minimum of $t_f^{(s)}$ and $t_f^{(m)}$. In 
addition, long-dashed horizontal lines show the damping time $t_d\simeq 1/\Gamma$ 
for increasing $\Gamma$ from top to bottom. Accordingly, for fixed $\gamma$ 
damping mechanisms play an increasing role with increasing $\Gamma$, 
cf.~text for details.}
\end{figure}

In the following analysis, we adopt a different notation and reformulate the expressions for the different time scales in terms of the fractional bremsstrahlung quantum energy $x=\omega/E$. Then, Eqs.~(\ref{equ:tfsEDomega}) and~(\ref{equ:tfmEDomega}) become 
\begin{eqnarray}
\label{equ:tfsED}
 t_f^{(s)} & \simeq & \frac{1}{xE-k(x)\beta} \,, \\
\label{equ:tfmED}
 t_f^{(m)} & \simeq & \sqrt{\frac{12E^2}{\hat{q} k(x)\beta}} 
\end{eqnarray}
with $k(x)=\sqrt{x^2E^2-m^2}$. The method of approximating $t_f$ by the minimum of $t_f^{(s)}$ and $t_f^{(m)}$ is sketched in Fig.~\ref{Fig:1new} using selected values for the 
entering quantities. The corresponding exact numerical solution of 
Eq.~(\ref{equ:X}) in terms of $x$ agrees with this estimate for small $x$ and 
for $x$ close to $1$, while it is typically somewhat smaller in the 
intermediate-$x$ region. 

\begin{figure}[t]
\centering
\vspace{5mm}
\includegraphics[width=0.45\textwidth]{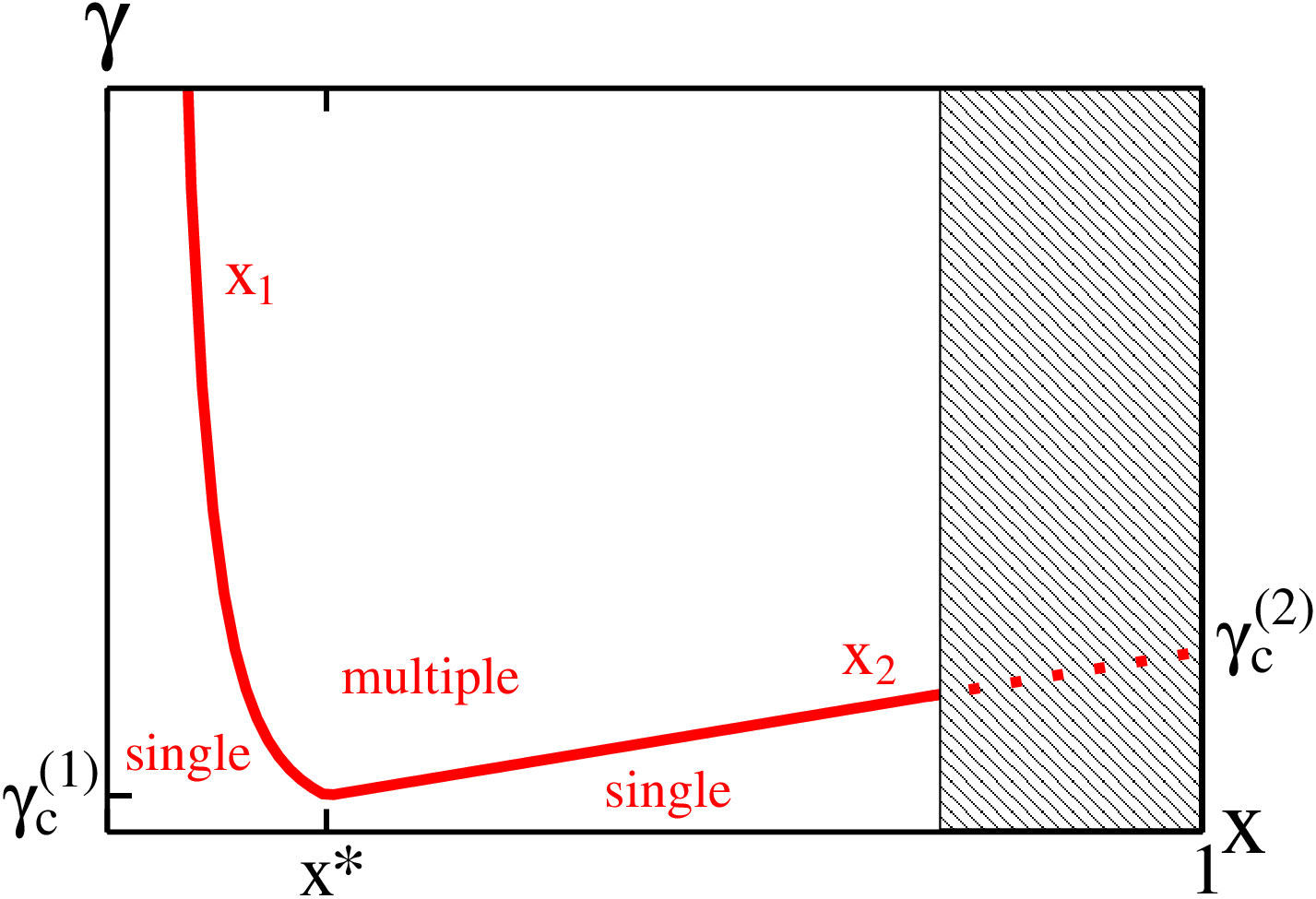}
\caption[]{\label{Fig:2new} (Color online) Sketch of the regions in 
$\gamma$-$x$-space, in which single or multiple scatterings will 
dominate the formation of radiation if damping effects are negligible using selected 
parameter values for visualization. While for $\gamma<\gamma_c^{(1)}$ 
the formation time is determined by $t_f^{(s)}$ for all $x$, for 
$\gamma>\gamma_c^{(1)}$ a region between $x_1$ and $x_2$ exists, 
in which $t_f^{(m)}$ determines 
$t_f$. The shaded region indicates that for a discussion of $x$ close to 
$1$ corrections become necessary. The depicted variables are summarized in Tab.~\ref{Tab:1}, 
cf.~also text for details.}
\end{figure}

Estimating the formation time by $\min\{t_f^{(s)}, t_f^{(m)}\}$ 
allows for a suitable analysis of the functional behaviour of $t_f$ with $x$ in 
dependence of $\gamma$ and of the parameter values for $M$, $m$ and $\hat{q}$. 
Moreover, in this way the interplay of $t_f$ with the damping time $t_d$ 
can be discussed easily. The results of the following analysis are 
summarized graphically in Figs.~\ref{Fig:1new},~\ref{Fig:2new} and~\ref{Fig:2newPLUS}, and the appearing variables and their definitions are collected in Tab.~\ref{Tab:1}. 

According to Eq.~(\ref{equ:X}) at leading order in $1/\gamma$ and 
for $m\ll xE$, one finds 
from Eq.~(\ref{equ:tfsED}) 
\begin{equation}
\label{equ:tfsEDapp}
 t_f^{(s)} \simeq \frac{2x\gamma M}{x^2M^2+m^2}
\end{equation}
while from Eq.~(\ref{equ:tfmED}) 
\begin{equation}
\label{equ:tfmEDapp}
 t_f^{(m)} \simeq \sqrt{\frac{12\gamma M}{\hat{q}x}} \,,
\end{equation}
follows. The function in Eq.~(\ref{equ:tfsEDapp}) exhibits a maximum at $x^*=m/M$ 
with $t_f^{(s)}(x^*)\simeq\gamma/m$ (cf.~Fig.~\ref{Fig:1new}). 
Moreover, it approaches $2/m$ for $\gamma\gg 1$ 
when $x$ approaches the minimal allowed value $x^*/\gamma$ which arises from 
the absence of radiation in the plasma for $xE<m$, cf.~\cite{Bluhm11}. 
We note that the exact expression for $t_f^{(s)}$ in Eq.~(\ref{equ:tfsED}) 
approaches $1/m$ instead. 

In the following, we restrict the discussion to the case $m\ll M$ such that $x^*\ll 1$. 
Then, one finds that $t_f^{(s)}(x)<t_f^{(m)}(x)$ over the whole $x$-range if 
$\gamma<\gamma_c^{(1)}\sim mM^2/\hat{q}$. In this case, the formation time will 
be given by $t_f^{(s)}$ only. For $\gamma > \gamma_c^{(1)}$, instead, one finds 
two values of $x$ as shown in Fig.~\ref{Fig:1new}, for which 
$t_f^{(s)}(x)=t_f^{(m)}(x)$. These intersection points are given by 
$x_1\sim x^*(\gamma_c^{(1)}/\gamma)^{1/3}$ for small $x$ and by 
$x_2\sim x^*\gamma/\gamma_c^{(1)}$ for larger $x$. We note that for 
$\gamma\to\gamma_c^{(1)}$ one finds $x_{1,2}\to x^*$. The formation times at the 
intersection points read $t_f(x_1)\simeq (\gamma/m)(\gamma_c^{(1)}/\gamma)^{1/3}$ 
and $t_f(x_2)=t_f^{\rm onset}$, where 
\begin{equation}
t_f^{\rm onset}\simeq\gamma_c^{(2)}/M 
\label{eq:def_tonset}
\end{equation} 
with $\gamma_c^{(2)}\sim M^3/\hat{q}$. 
The scale $t_f^{\rm onset}$ is independent of $\gamma$ and represents the minimal amount of time 
necessary to radiate a photon through a multiple scattering process. 
To obtain the intersection points and the 
corresponding formation times, we have approximated $t_f^{(s)}$ in 
Eq.~(\ref{equ:tfsEDapp}) either by $t_f^{(s)}\simeq 2x\gamma M/m^2$ for small $x\ll x^*$, 
cf.~also~\cite{Kampfer00}, or by $t_f^{(s)}\simeq 2\gamma/(xM)\simeq t_{BH}$ for larger 
$x\gg x^*$. 

Between $x_1$ and $x_2$, $t_f^{(s)}(x)>t_f^{(m)}(x)$ and therefore 
$t_f^{(m)}$ determines the formation time of radiation, whereas outside of this region 
radiation is still dominated by single scatterings, in particular for 
$x<x_1$, cf.~also~\cite{Galitsky}. 
The behaviour observed for $m\ne 0$ is in striking contrast 
to the $m=0$ case. In this case, the expressions for the formation times become $t_f^{(s)}\to t_{BH}\simeq 2\gamma^2/(xE)$ and 
$t_f^{(m)}\to\sqrt{12E/(\hat{q}x)}$, which are equivalent to the 
known results~\cite{Landau53} for bremsstrahlung photons from incoherent 
(single) and coherent (multiple) scatterings, respectively. In the region $0\leq x\lesssim x_2$ one finds $t_f^{(s)}(x)>t_f^{(m)}(x)$ such that $t_f$ is also for small 
$x$ determined by $t_f^{(m)}$, cf.~\cite{Landau53}. The 
difference stems from the evident reduction of $t_f^{(s)}$ compared to $t_{BH}$ 
in the small-$x$ region due to polarization effects when $m\ne 0$. 

The lower bound $x_1$ decreases with $\gamma^{-1/3}$, while the upper bound 
$x_2$ increases linearly with $\gamma$. Consequently, the region in which 
multiple scattering processes become effective increases with increasing $\gamma$ 
(or increasing energy $E$ for fixed $M$) as 
evident from Fig.~\ref{Fig:2new}. For $\gamma\to\gamma_c^{(2)}$, 
$x_2$ tends to $1$. Thus, for $\gamma>\gamma_c^{(2)}$ one would expect to find only one intersection point, for which 
$t_f^{(s)}(x)=t_f^{(m)}(x)$, so that for $x\sim 1$ the BH-spectrum in Eq.~(\ref{equ:3}) 
could not be recovered, cf.~\cite{BluhmQM11}. However, with 
increasing $x$ towards $1$ corrections, which are not present in 
Eq.~(\ref{equ:X}), have to be taken into account. These are 
discussed later in Sec.~\ref{sec:4_1} in the case of QCD. 
Generically, such corrections lead to a reduction of both $t_f^{(s)}$ and 
$t_f^{(m)}$ for $x$ close to $1$, where $t_f^{(s)}$ becomes stronger reduced. 
This implies that the region in which multiple scattering processes dominate 
effectively shrinks and the Bethe-Heitler bremsstrahlung spectrum is 
necessarily recovered for $x\sim 1$. As long as $m\ll M$ and 
$\gamma<\gamma_c^{(2)}$, one may refrain from including corrections for $x$ 
close to $1$. Otherwise, these corrections would become mandatory, which is, 
however, beyond the scope of this section. 

\subsection{Competition between photon formation and damping \label{sec:3_2}}

In an absorptive medium, damping of radiation effects can modify the 
aforementioned picture significantly. Considering $m\ll xE$ and 
$\Gamma\ll xE$, the exponential damping factor in Eq.~(\ref{equ:DampingFactor}) becomes 
$\mathcal{F}(\bar{t}\,)\simeq e^{-\Gamma \bar{t}}$. This gives rise to a damping time, 
$t_d\simeq 1/\Gamma$ (cf.~Eq.~(\ref{equ:DampingTime})), as highlighted in Fig.~\ref{Fig:1new} by long-dashed horizontal lines, at which, compared to the undamped case, the amplitude in Eq.~(\ref{equ:1}) is reduced by a factor $1/e$. 

\begin{figure*}[ht]
\centering
\vspace{5mm}
\includegraphics[width=0.45\textwidth]{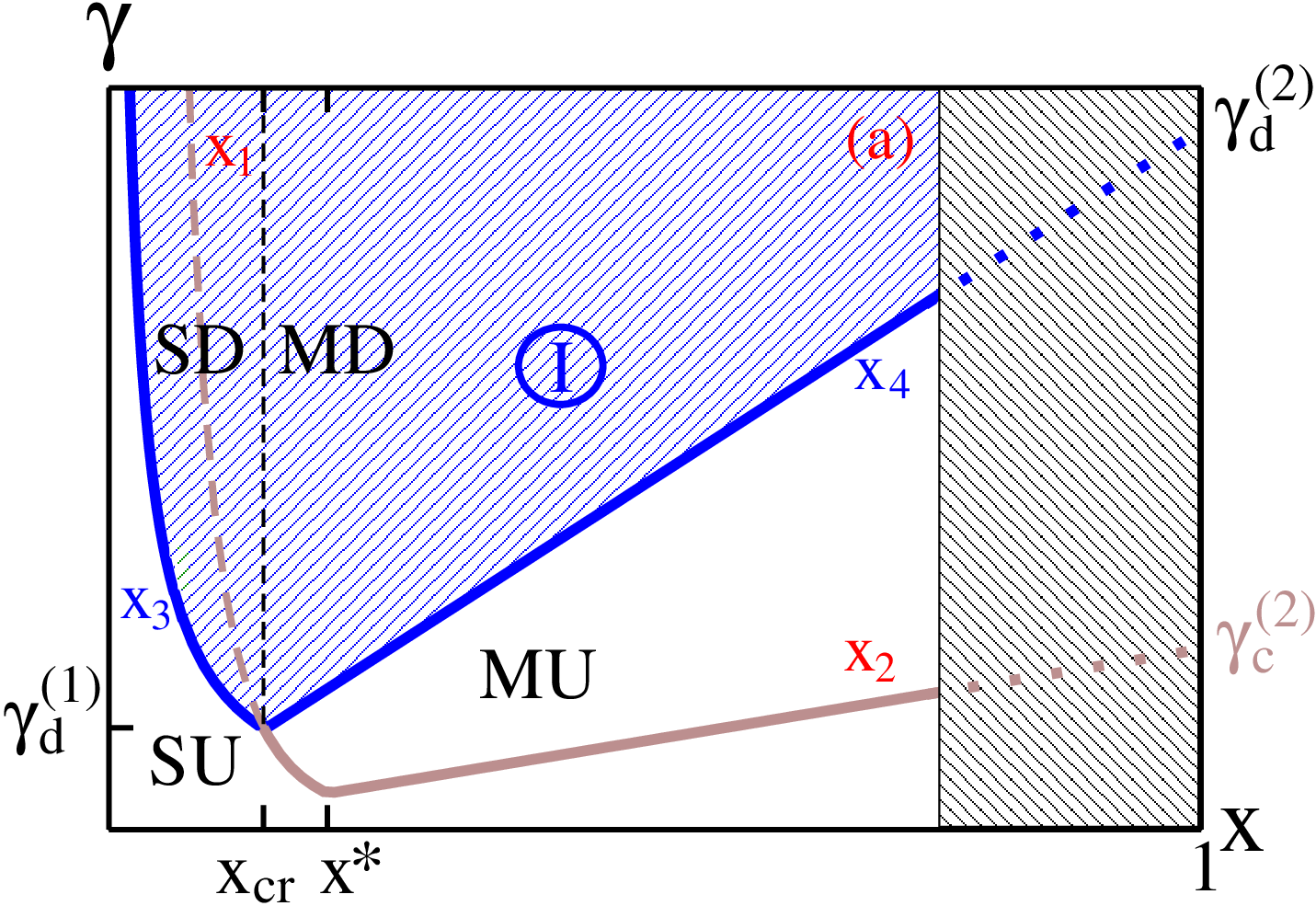} 
\hspace{3mm}
\includegraphics[width=0.45\textwidth]{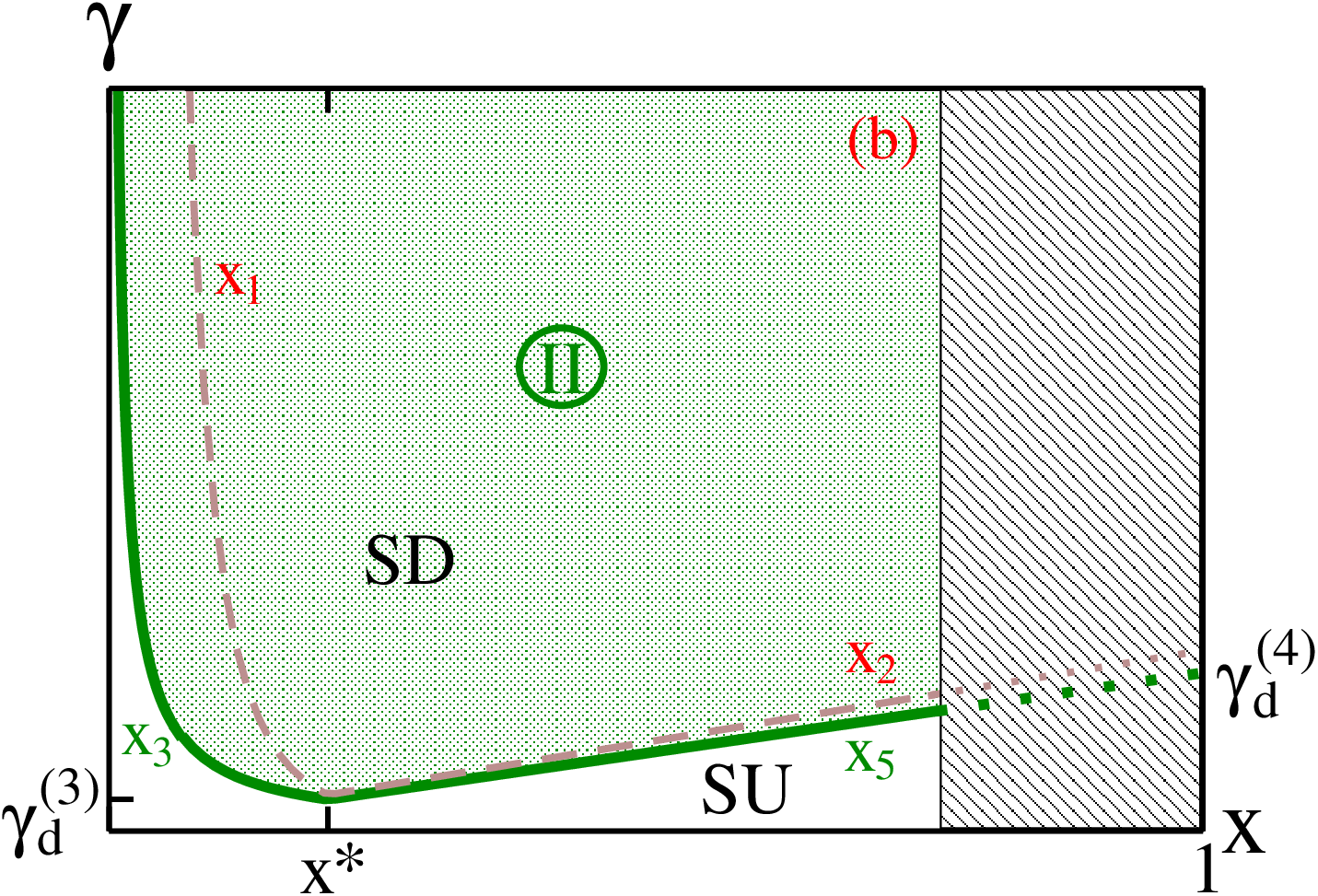} 
\caption[]{\label{Fig:2newPLUS} (Color online) (a): Sketch of the region (shaded region (I)) in $\gamma$-$x$-space, which is influenced by damping effects for fixed $\Gamma<\hat{q}/M^2$. As evident, for a fixed $\Gamma$ damping plays an increasing role with 
increasing $\gamma$. (b): As in panel (a) but for fixed $\Gamma>\hat{q}/M^2$ (shaded region (II)). The acronyms SU, MU, MD and SD stand for single undamped, multiple undamped, multiple damped and single damped, respectively, as explained in detail in Sec.~\ref{sec:3_3}, and $x_{cr}\sim\sqrt{m^2\Gamma/(M^2\Gamma-\hat{q})}$. The other depicted variables are summarized in Tab.~\ref{Tab:1}, cf.~also text for details.}
\end{figure*} 
In case $t_d$ is large compared to the formation time of radiation, damping mechanisms will not significantly 
affect the spectrum, as evident from Fig.~\ref{Fig:0new}. This situation is also illustrated by the upper long-dashed horizontal line in Fig.~\ref{Fig:1new}. If, in contrast, $t_d\lesssim t_f$ in 
a specific region in $x$, then damping will be of significance in that region. 
In particular, it implies that any coherent interference of 
radiation amplitudes happens only during $t_d$ rather than $t_f$, 
because the formation process of the radiation quantum will be hampered by damping mechanisms. Strictly speaking, any finite value of $\Gamma$ enters $\vert n_r\vert$ and 
thus modifies Eq.~(\ref{equ:X}). The above parametric 
discussion of $t_f$, however, is still valid qualitatively as long as we 
demand $\Gamma<m$. 

The relevant scales for discussing the influence of damping mechanisms are 
$t_d$ and $t_f^{\rm onset}\simeq M^2/\hat{q}$ or, equivalently, 
$\Gamma$ and $\hat{q}/M^2$. In case $t_d>t_f^{\rm onset}$ or $\Gamma<\hat{q}/M^2$, 
one expects to find a region in $\gamma$-$x$-space, in which coherence effects are of importance. 
In the opposite case, however, i.e.~if $t_d<t_f^{\rm onset}$ or $\Gamma>\hat{q}/M^2$, 
damping effects will be significant and coherence effects will play no role. 
This stems from the definition of $t_f^{\rm onset}$, see the comment below 
Eq.~(\ref{eq:def_tonset}). 
Again, we do not include corrections necessary for $x$ close to $1$ in this 
discussion. 

\begin{table*}
\vspace{5mm}
\centering
 \begin{tabular}[t]{|l |l|l|}
  \hline
  \rule{0pt}{2ex}
  $\gamma$-scales & noticeable points 
  & conditions\\
  \hline
  \rule{0pt}{3ex}
  $\gamma_c^{(1)}\sim mM^2/\hat{q}$ & $x^*=m/M$ & \\
  $\gamma_c^{(2)}\sim M^3/\hat{q}$ &  $x_1\sim x^*(\gamma_c^{(1)}/\gamma)^{1/3}\sim (m^4/(\gamma M\hat{q}))^{1/3}$ 
    & $t_f^{(s)}(x_1)=t_f^{(m)}(x_1),\,\, x\ll x^*$ \\
  $\gamma_d^{(1)}\sim \sqrt{\hat{q}m^2/(\Gamma^3 M^2)}$ & 
    $x_2\sim x^*\gamma/\gamma_c^{(1)}\sim \gamma/\gamma_c^{(2)}\sim \gamma\hat{q}/M^3$ & 
    $t_f^{(s)}(x_2)=t_f^{(m)}(x_2),\,\, x\gg x^*$ \\
  $\gamma_d^{(2)}\sim \hat{q}/(\Gamma^2 M)$ &  
    $x_3\sim x^*\gamma_d^{(3)}/\gamma\,\sim m^2/(\gamma M\Gamma)$ & 
    $t_d=t_f^{(s)}(x_3),\,\, x\ll x^*$ \\
  $\gamma_d^{(3)}\sim m/\Gamma$ &  $x_4\sim \gamma/\gamma_d^{(2)}\sim \gamma M\Gamma^2/\hat{q}$ 
    & $t_d=t_f^{(m)}(x_4)$  \\
  $\gamma_d^{(4)}\sim M/\Gamma$ & $x_5\sim \gamma/\gamma_d^{(4)}\sim\gamma\Gamma/M$ & 
    $t_d=t_f^{(s)}(x_5),\,\, x\gg x^*$ \\
  \hline
 \end{tabular}
 \caption[]{\label{Tab:1} Summary of the definitions of the variables appearing in the text and depicted in Figs.~\ref{Fig:1new},~\ref{Fig:2new} and~\ref{Fig:2newPLUS} as determined from these definitions.}
\end{table*}

Considering first damping rates $\Gamma<\hat{q}/M^2$ (case sketched in panel (a) of 
Fig.~\ref{Fig:2newPLUS}), one finds that damping plays a negligible role for 
$\gamma<\gamma_d^{(1)}\sim(m/\Gamma)\sqrt{\hat{q}/(\Gamma M^2)}$. For such 
$\gamma$, $t_d>t_f(x_1)$ and, thus, $t_d$ is larger than $t_f$ for any $x$. 
It implies, in particular, that for $\gamma<\gamma_c^{(1)}$ undamped single 
scatterings dominate the formation of radiation as before because 
$\gamma_c^{(1)}<\gamma_d^{(1)}$ by definition. For $\gamma>\gamma_d^{(1)}$, 
instead, damping mechanisms become relevant in a region between $x_3$ and 
$x_4$ (region (I) in panel (a) of Fig.~\ref{Fig:2newPLUS}). 
These boundary points are determined from the conditions 
$t_d=t_f^{(s)}(x_3)$ for $x_3\ll x^*$ and $t_d=t_f^{(m)}(x_4)$ 
(leading to $x_4>x_1$), respectively. They read 
$x_3\sim x^*m/(\Gamma\gamma)\sim x^*(\gamma_d^{(1)}/\gamma)\sqrt{\Gamma M^2/\hat{q}}$ 
and $x_4\sim\Gamma^2\gamma M/\hat{q}$, cf.~also~\cite{Galitsky}. 
This situation is also illustrated by the middle long-dashed horizontal line in 
Fig.~\ref{Fig:1new}. 

The lower bound $x_3\propto\gamma^{-1}$ decreases faster than $x_1$ with increasing 
$\gamma$ but tends to $x_1$ as $\gamma\to\gamma_d^{(1)}$, while the upper 
bound $x_4\propto\gamma$ increases slower than $x_2$ with increasing $\gamma$ and tends to $1$ as 
$\gamma\to\gamma_d^{(2)}\sim\hat{q}/(\Gamma^2M)$. This leads to the situation 
exhibited in Fig.~\ref{Fig:2newPLUS} (panel (a)). For $\gamma>\gamma_d^{(2)}$, damping mechanisms 
are significant for the whole $x$-range above $x_3$, while for $\gamma_d^{(2)}>\gamma>\gamma_d^{(1)}$ 
a region for $x>x_4$ emerges, in which 
undamped multiple scattering processes become relevant. 
In case $\gamma_c^{(2)}>\gamma>\gamma_d^{(1)}$, this picture will be modified 
to the extent that, in addition, a region for $x>x_2(>x_4)$ exists, in which 
undamped single scatterings dominate the formation of radiation again. 

In the special case $\Gamma\to\hat{q}/M^2$, one finds that 
$\gamma_d^{(1)}\to\gamma_c^{(1)}$ and $\gamma_d^{(2)}\to\gamma_c^{(2)}$, where 
$x_4\to x_2$. This implies, in particular, that for any 
$\gamma>\gamma_c^{(1)}$ coherence effects are superseded by damping mechanisms and, thus, play no significant role for the formation of radiation anymore. 

For even larger damping rates $\Gamma>\hat{q}/M^2$ (case sketched in panel (b) of Fig.~\ref{Fig:2newPLUS}), damping effects will be influential already for $\gamma<\gamma_c^{(1)}$: As $t_f$ is given by $t_f^{(s)}$ for 
all $x$ in this case, another scale 
$\gamma_d^{(3)}\sim m/\Gamma<\gamma_c^{(1)}$ arises from assuming 
that $t_d=t_f^{(s)}(x^*)$. Accordingly, damping mechanisms become important 
for $\gamma>\gamma_d^{(3)}$ in a region between $x_3$ and 
$x_5\sim\Gamma\gamma/M$ (region (II) in panel (b) of Fig.~\ref{Fig:2newPLUS}), while outside of this region 
radiation is still formed by undamped single scatterings. The boundary point $x_5$ is determined 
from the condition $t_d=t_f^{(s)}(x_5)$ for $x_5\gg x^*$. This picture does not 
change for $\gamma>\gamma_c^{(1)}$. Because already $t_d<t_f^{\rm onset}=t_f(x_2)$, damping 
effects are also in this case important in the region between $x_3$ and 
$x_5>x_2$ (as also illustrated by the lower long-dashed horizontal line in 
Fig.~\ref{Fig:1new}). Therefore, damping mechanisms hamper the formation of radiation
in a large region of 
$\gamma$-$x$-space for $\Gamma>\hat{q}/M^2$. We note that $x_5$ tends to $1$ as 
$\gamma\to\gamma_d^{(4)}\sim M/\Gamma$. 

\subsection{Physical implications \label{sec:3_3}}

The above analysis discussed the competition of the time scales $t_f^{(s)}$, $t_f^{(m)}$ and $t_d$. 
In this section, we want to interpret our findings in the context of the spectrum. In case of negligible damping, i.e.~$t_d$ large compared to the other time scales, the spectrum behaves in line with Eq.~(\ref{eq:small_damping}). In case of a non-negligible damping rate, however, we identified three different regimes, in which, depending on $x$ and $\gamma$, damping effects can be important: (a) when $t_d\ll t_f^{(s)}\ll t_f^{(m)}$, (b) when $t_d\ll t_f^{(m)\,2}/t_f^{(s)}\ll t_f^{(m)}\ll t_f^{(s)}$, and (c) when $t_f^{(m)\,2}/t_f^{(s)}\ll t_d\ll t_f^{(m)}\ll t_f^{(s)}$. 

Among these regimes, regime (b) is particularly interesting. From our discussion of the competition of the different time scales in Secs.~\ref{sec:3_1} and~\ref{sec:3_2}, one could understand the physical process taking place as the hampering of the formation of radiation in a multiple scattering process, because without damping in the medium radiation would be formed coherently ($t_f^{(m)}\ll t_f^{(s)}$). In view of the spectra, however, the regimes (a), in which $t_f^{(s)}\ll t_f^{(m)}$, and (b) yield equivalent results, cf. the discussion around Eq.~(\ref{eq:big_damping_small_ts}). This observation can be understood by interpreting that in the case $t_d\ll t_f^{(m)\,2}/t_f^{(s)}$, the influence of damping effects is so strong that effectively the radiation formation is hampered already after a single scattering of the charge in the absorptive medium even if $t_f^{(m)}\ll t_f^{(s)}$ in the absence of damping. 

This particular physical situation is realized in parts of the regions (SD) highlighted in Fig.~\ref{Fig:2newPLUS}, in which single damped scatterings dominate the physics. More precisely, the conditions for regime (b) are met in the shaded region between $x_1$ and $x_{cr}\sim \sqrt{m^2\Gamma/(M^2\Gamma-\hat{q})}$ in panel (a) of Fig.~\ref{Fig:2newPLUS} and between $x_1$ and $x_2$ in panel (b) of Fig.~\ref{Fig:2newPLUS}. Here, $x_{cr}$ represents a $\gamma$-independent critical $x$-value determined from assuming $t_d=t_f^{(m)\,2}/t_f^{(s)}$ which, in case of $\Gamma<\hat{q}/M^2$, allows for a discrimination of the nature of the physical process taking place in the $\gamma$-$x$-space regions, in which damping plays a role (i.e. in region (I) of Fig.~\ref{Fig:2newPLUS} panel (a)). With increasing $\Gamma\to\hat{q}/M^2$, one finds $x_{cr}\to\infty$.

Corresponding to this interpretation, the physical picture simplifies tremendously: For $\Gamma<\hat{q}/M^2$ (see panel (a) of Fig.~\ref{Fig:2newPLUS}) and any fixed $x<x_{cr}$, damping effects are strong enough to influence the radiation formation significantly already after a single scattering of the charge (shaded region (SD) between $x_3$ and $x_{cr}$), while for fixed $x>x_{cr}$ damping effects influence the formation of radiation in multiple scattering processes only (shaded region (MD) between $x_{cr}$ and $x_4$), provided $\gamma$ is large enough. For $x<x_{cr}$, one passes therefore with increasing $\gamma$ from single undamped (region (SU) below the $x_3$-curve) to single damped (shaded region (SD)) scattering processes, which determine the 
spectrum. For $x>x_{cr}$, instead, one passes from single undamped (region (SU) below the $x_1$-$x_2$-curve) to multiple undamped (region (MU) in between the $x_4$-curve and the $x_1$-$x_2$-curve) to multiple damped (shaded region (MD)) scatterings with increasing $\gamma$. 

For $\Gamma\geq\hat{q}/M^2$ (see panel (b) of Fig.~\ref{Fig:2newPLUS}), damping effects are so strong that effectively the radiation formation is hampered for any $x$ after a single scattering of the charge (shaded region (SD) between $x_3$ and $x_5$) as long as $\gamma$ is large enough. Otherwise, single undamped scatterings determine the spectrum (region (SU) below the $x_3$-$x_5$-curve), and there is no room left for coherence effects. 

\section{Application to QCD phenomenology \label{sec:4}}

We now extend our study to the consideration of damping phenomena in a 
strongly interacting medium. To our best knowledge, the 
influence of the damping of radiation has not yet been investigated in the 
context of radiative energy loss in QCD matter. Given the universality of the possible physical processes, it is reasonable to assume that dissipative effects lead equally to a damping of the radiation spectrum in QCD matter, which may formally be written down similar to Eq.~(\ref{eq:spectrum_ratio}) up to a different normalization factor and essential differences in the entering time scales due to the non-Abelian character of QCD. 
In Sec.~\ref{sec:4_1}, we start our discussion, as in Sec.~\ref{sec:3_1}, by analyzing the functional behaviour and the parametric 
dependence of the formation time $t_f$ of gluon bremsstrahlung by ignoring 
possible damping mechanisms. In this case, estimates for $t_f$ exist, 
cf.~e.g.~\cite{Arnold08}. Then, in Sec.~\ref{sec:4_2}, we include damping phenomena into the 
discussion in the same way as done in Sec.~\ref{sec:3_2} by introducing a 
competing time scale $t_d$. The phenomenological consequences for the gluon bremsstrahlung spectrum are analyzed in Sec.~\ref{sec:4_4} in analogy to Sec.~\ref{sec:3_3} and a discussion of possible physical mechanisms 
leading to the damping of the gluon radiation is found in Sec.~\ref{sec:4_3}.

\subsection{Analysis of the gluon bremsstrahlung formation time in the absence of damping \label{sec:4_1}}

The formation time of gluon bremsstrahlung in the QCD plasma can be 
estimated, following~\cite{Arnold08}, from the inverse expectation value of 
the energy imbalance between the final state (a relativistic on-mass-shell 
parton with momentum $\vec{p}$ plus an on-mass-shell bremsstrahlung gluon 
with momentum $\vec{k}$) and a relativistic on-mass-shell single parton 
state carrying the same total momentum $\vec{P}=\vec{p}+\vec{k}$. This leads 
to the condition~\cite{Arnold08} 
\begin{equation}
\label{equ:formtimeQCD}
 t_f \left[\frac{\langle p_B^2\rangle+x^2 m_s^2 + (1-x) m_g^2}{2x(1-x)E}
 \right] \simeq 1 \,,
\end{equation}
where $m_s$ and $m_g$ denote the masses of the emitting color charge and of 
the emitted bremsstrahlung gluon, respectively, while $\vec{p}_B$ is defined 
as $\vec{p}_B=(p\vec{k}_\perp-k\vec{p}_\perp)/P$ and $E\simeq P$. 

In order to make a connection with Sec.~\ref{sec:3_1}, we consider in 
Eq.~(\ref{equ:formtimeQCD}) first the special case in which 
$\vec{k}_\perp$ vanishes, while the emitting charge experiences small 
deflections through soft, elastic scatterings in the medium. Accordingly, 
$\langle p_B^2\rangle\simeq x^2t_f\hat{q}_s$, where $\hat{q}_s$ is the mean 
squared transverse momentum per unit time picked up by the charge, and 
Eq.~(\ref{equ:formtimeQCD}) renders into 
\begin{equation}
\label{equ:formtimeQED}
 t_f \frac{x}{2E} \left[\frac{m_s^2}{(1-x)}+\frac{m_g^2}{x^2}\right]+
 t_f^2 \frac{x\hat{q}_s}{2E(1-x)} \simeq 1 \,.
\end{equation}
It is interesting to note that in the small $x\ll 1$ limit this 
condition equation will be equivalent to the condition Eq.~(\ref{equ:X}) discussed in detail above, apart from a factor 
$\hat{q}=6\hat{q}_s$, if in Eq.~(\ref{equ:X}), now written in terms of $x$, the function $k(x)$ 
is expanded for $m\ll xE$, only terms of $\mathcal{O}(1/\gamma)$ are kept, and 
$M$ and $m$ are identified with $m_s$ and $m_g$, respectively. 

The dominant contribution to the radiative energy loss of a relativistic 
parton in the hot QCD plasma stems, however, from the rescatterings of the 
radiated gluon in the medium~\cite{Baier95}. Thus, we consider in 
Eq.~(\ref{equ:formtimeQCD}) rather the case in which the bremsstrahlung 
gluon undergoes soft, elastic scatterings during its formation, while any 
deflection of the emitting color charge is assumed to be of minor 
importance. This leads to 
$\langle p_B^2\rangle\simeq (1-x)^2\langle k_\perp^2\rangle\simeq 
(1-x)^2 t_f\hat{q}_g$, where $\hat{q}_g$ 
is the mean squared transverse momentum per unit time picked up by the 
radiated gluon. Then, the condition for the gluon formation time 
Eq.~(\ref{equ:formtimeQCD}) becomes 
\begin{equation}
\label{equ:formtimeQCDglue}
 t_f \frac{\left[ x^2m_s^2+m_g^2(1-x)\right]}{2x(1-x)E} + 
 t_f^2 \frac{(1-x)\hat{q}_g}{2xE} \simeq 1 \,.
\end{equation}

We note that the above equation is strictly applicable only in the limit 
of highly energetic gluons, i.e.~when $k_{\|}\simeq xE$. For gluon energies 
of the order of the gluon mass, in contrast, it has to be generalized. 
This can be achieved conveniently by considering rather the imbalance of 
parallel momenta in the above mentioned scattering process, while the energy is 
conserved. The resulting condition equation for the gluon formation length (the formation length translates easily into a formation time) 
looks similar to Eq.~(\ref{equ:formtimeQCDglue}), where one has to replace 
only $m_g^2/(2xE)$ by $m_g^2/(\omega+k_{\|})$ and $(1-x)\hat{q}_g$ by 
$[2\omega/(\omega+k_{\|})-x]\hat{q}_g$. As a self-consistency condition, the 
gluon energy must satisfy 
\begin{equation}
\omega\geq\sqrt{m_g^2+\langle k_\perp^2\rangle} \,. 
\label{eq:SCC}
\end{equation}
These modifications introduce, however, merely numerical changes of 
order $\mathcal{O}(1)$ in the prefactors so that Eq.~(\ref{equ:formtimeQCDglue}) 
is completely sufficient for the parametric discussion we aim at in the following. 

\begin{figure}[t]
\centering
\vspace{1.5mm}
\includegraphics[width=0.45\textwidth]{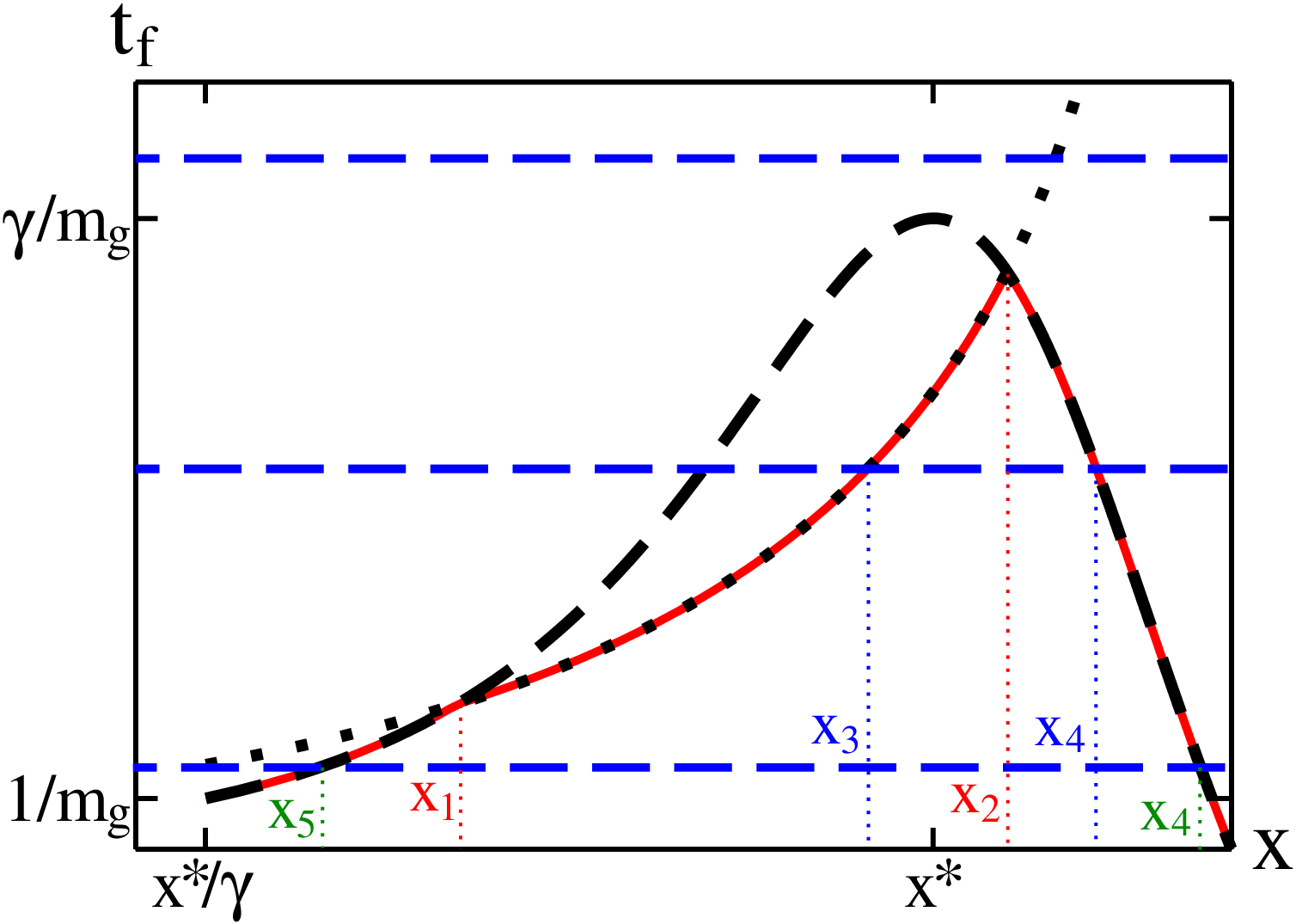} 
\caption[]{\label{Fig:4new} (Color online) Visualization of the formation time 
$t_f$ for gluon bremsstrahlung as a function of $x$ for selected parameter values and fixed $\gamma$. 
Short-dashed and dotted 
curves exhibit $t_f^{(s)}$ and $t_f^{(m)}$, respectively, as defined in 
Eqs.~(\ref{equ:tfsQCD}) and~(\ref{equ:tfmQCD}), while the solid curve depicts 
our estimate for $t_f$ given by the minimum of $t_f^{(s)}$ and $t_f^{(m)}$. Moreover, long-dashed 
horizontal lines indicate a fixed damping time $t_d$, where damping 
increases from top to bottom, cf.~text for details.}
\end{figure}

\begin{figure*}[t]
\centering
\vspace{5mm}
\includegraphics[width=0.45\textwidth]{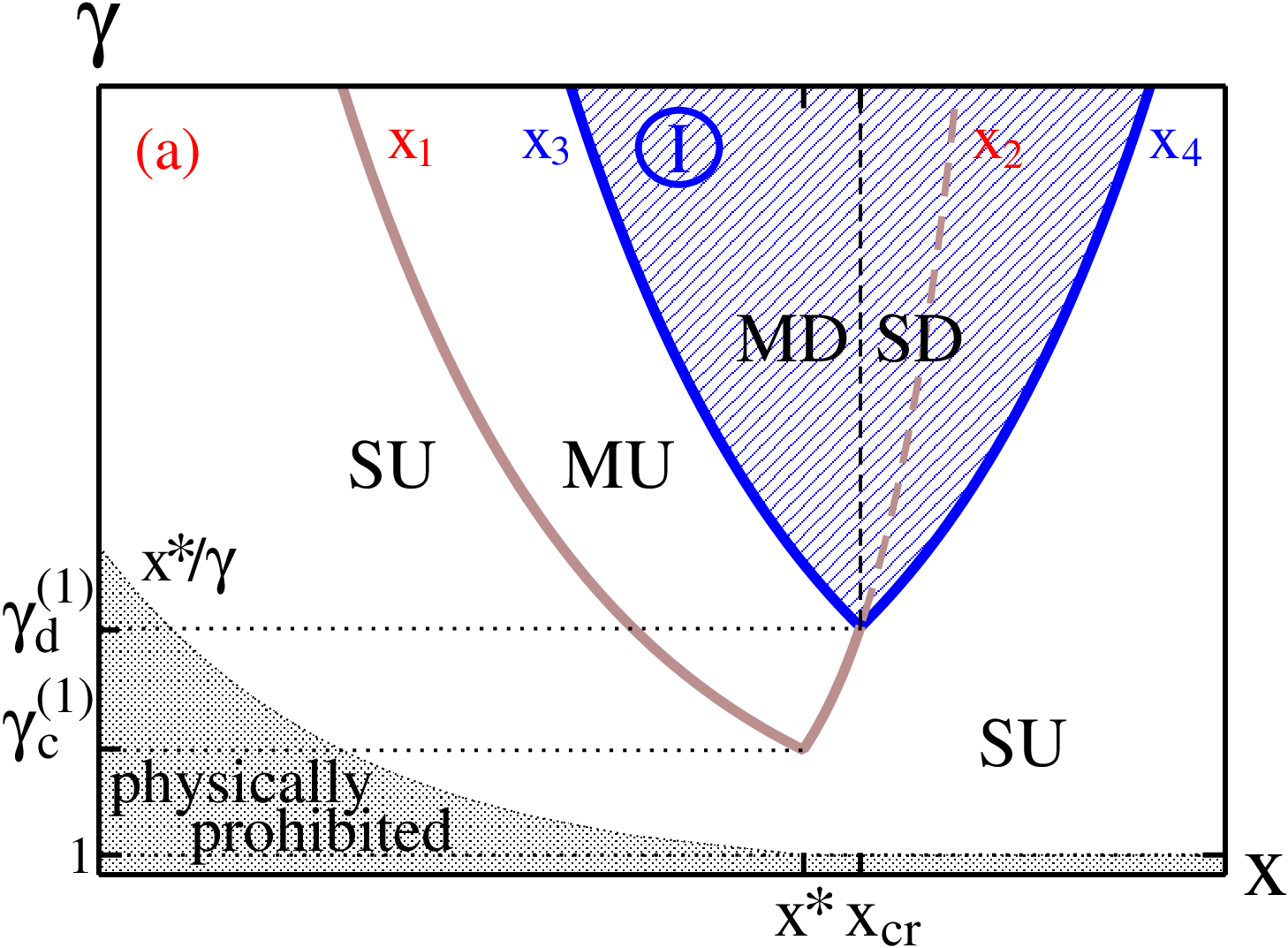} 
\hspace{3mm}
\includegraphics[width=0.45\textwidth]{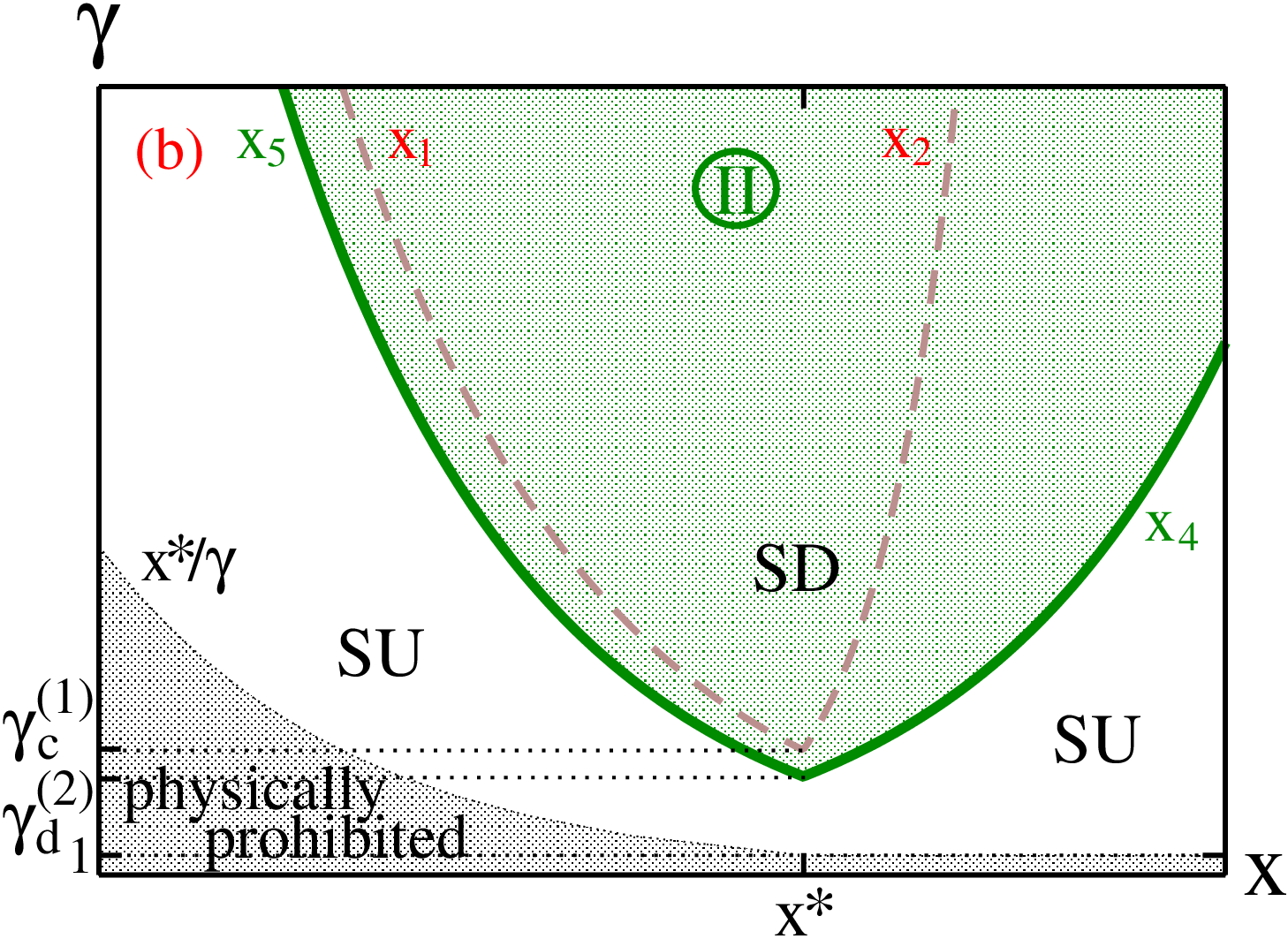} 
\caption[]{\label{Fig:5new} (Color online) Sketch of the regions in $\gamma$-$x$-space, in which different physical processes dominate the formation of gluon bremsstrahlung. Here, selected parameter values are used, for which $\hat{q}_g<m_g^3$. The physically prohibited regions correspond to values incompatible with the self-consistency condition Eq.~(\ref{eq:SCC}). In case damping effects are negligible, coherence effects are of importance in the region between the curves $x_1$ and $x_2$, while outside of this region single scatterings dominate the formation of radiation. In case of non-negligible damping, this picture is altered and damping mechanisms become important in the shaded regions: For fixed $\Gamma<\hat{q}_g/m_g^2$ (case shown in panel (a)), damping effects play a role in region (I), while for fixed $\Gamma>\hat{q}_g/m_g^2$ (case shown in panel (b)) they are important in region (II). The acronyms SU, MU, MD and SD are explained in Sec.~\ref{sec:4_4} and have the same physical meaning as in the QED case (see also Fig.~\ref{Fig:2newPLUS}), while the $\gamma$-independent quantity 
$x_{cr}$ separating the regions (MD) and (SD) is determined from the condition $t_d=t_f^{(m)\,2}/t_f^{(s)}$. The other depicted variables are summarized in Tab.~\ref{Tab:2}, cf. also text for details. We note that $x_2$ and $x_4$ have been approximated by $x_2\sim C$ and $x_4\sim\gamma\Gamma/m_s$.}
\end{figure*}

\begin{figure}[t]
\centering
\vspace{1.5mm}
\includegraphics[width=0.45\textwidth]{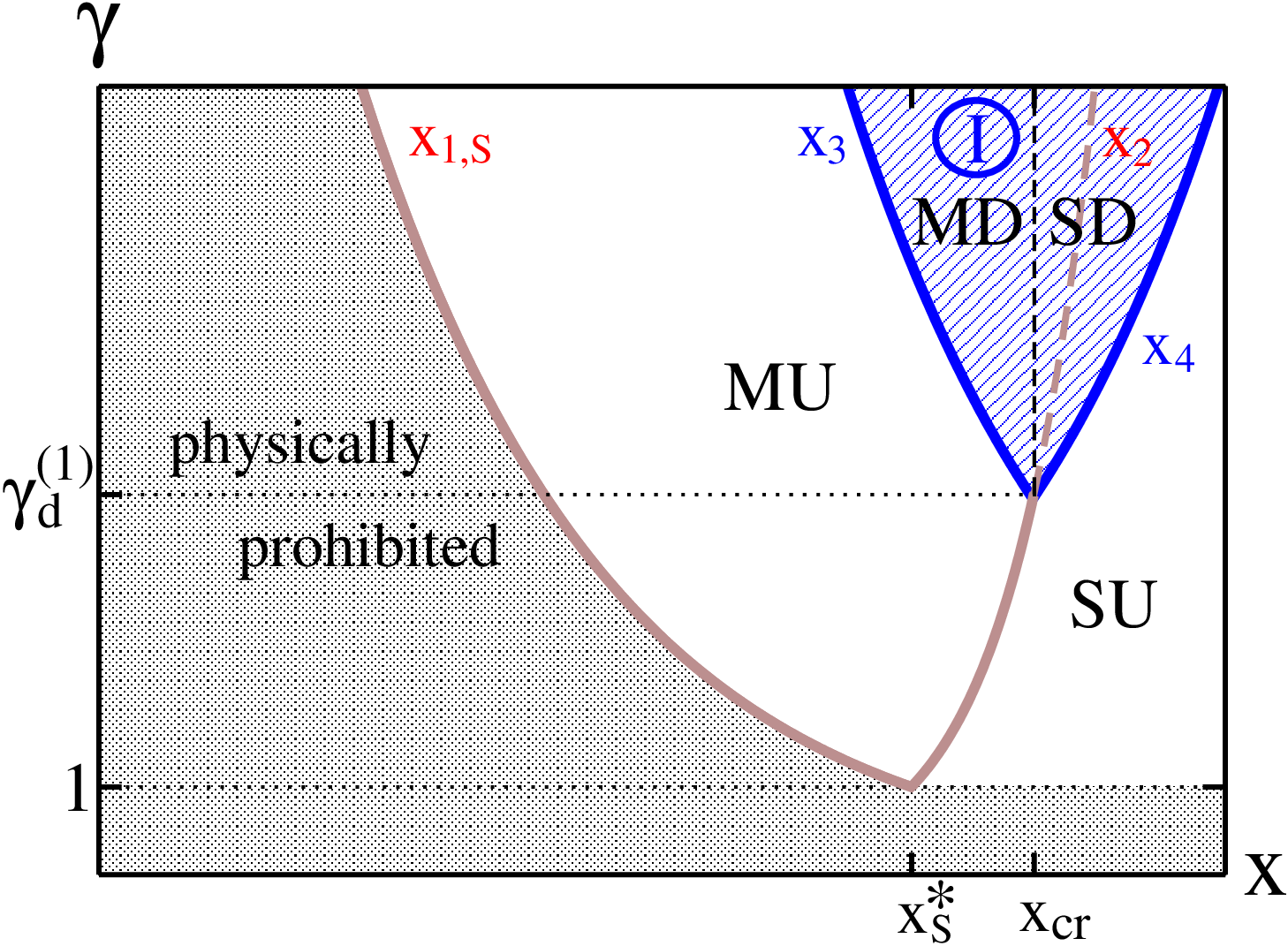} 
\caption[]{\label{Fig:5+1new} (Color online) Similar to Fig.~\ref{Fig:5new}, but with parameter values for which $m_g^3<\hat{q}_g$. 
The physically prohibited region corresponds to values incompatible with the self-consistency condition Eq.~(\ref{eq:SCC}). 
In case damping effects are negligible, coherence effects are important for any $\gamma$ in the region between $x_{1,S}$ and $x_2$, where $x_{1,S}$ is the minimal allowed $x$-value. Outside of this region, single scatterings dominate the formation of radiation. For non-negligible $\Gamma<\hat{q}_g/m_g^2$, this picture changes and damping effects play a significant role in the shaded region (I), cf.~text for details.}
\end{figure}

As in Sec.~\ref{sec:3_1}, the functional behaviour of the formation time 
for gluon bremsstrahlung $t_f$ with $x$ depending on $\gamma=E/m_s$ and on the 
parameter values for $m_s$, $m_g$ and $\hat{q}_g$ can be analyzed 
conveniently by estimating $t_f$ by the minimum of $t_f^{(s)}$ and 
$t_f^{(m)}$ defined as 
\begin{eqnarray}
\label{equ:tfsQCD}
 t_f^{(s)} & \simeq & \frac{2x(1-x)\gamma m_s}{x^2m_s^2+m_g^2(1-x)} \,, \\
\label{equ:tfmQCD}
 t_f^{(m)} & \simeq & \sqrt{\frac{2x\gamma m_s}{(1-x)\hat{q}_g}} \,,
\end{eqnarray}
which both increase with $\gamma$, $t_f^{(s)}$ showing the stronger dependence. 
Similar to the previous sections, $t_f^{(s)}$ 
denotes the gluon formation time in incoherent (single) scatterings, while 
$t_f^{(m)}$ represents the formation time of bremsstrahlung gluons in a 
coherent (multiple) scattering process. This procedure for determining 
$t_f$ is shown in Fig.~\ref{Fig:4new}. We stress here that the above 
expression for $t_f^{(s)}$ is only applicable for finite $m_g$, while in 
the limit of a vanishing in-medium gluon mass it has to be modified as is 
discussed below. The results of the following analysis are summarized 
graphically in Figs.~\ref{Fig:4new}, \ref{Fig:5new} and~\ref{Fig:5+1new}, while the 
definitions of the appearing variables are listed in Tab.~\ref{Tab:2}. 

Compared to Eqs.~(\ref{equ:tfmEDapp}) and~(\ref{equ:tfsEDapp}), corrections 
for $x$ close to $1$ are now taken into account in both expressions for 
$t_f^{(s)}$ and $t_f^{(m)}$. These modify the behaviour of the time scales for $x\sim 1$. 
Apart from these modifications, $t_f^{(s)}$ is rather similar to the expression in Sec.~\ref{sec:3_1}, while 
the functional form of $t_f^{(m)}$ is significantly different. The function 
$t_f^{(s)}$ in Eq.~(\ref{equ:tfsQCD}) exhibits a maximum at $m_g/(m_s+m_g)$. In the following, we focus our 
discussion on the case $m_g\ll m_s$, such that the position of this 
maximum becomes $x^*=m_g/m_s$ with $t_f^{(s)}(x^*)\simeq\gamma/m_g$ (cf.~Fig.~\ref{Fig:4new}) similar 
to Sec.~\ref{sec:3_1}. Moreover, $t_f^{(s)}$ from 
Eq.~(\ref{equ:tfsQCD}) may be approximated by $t_f^{(s)}\simeq 2x\gamma m_s/m_g^2$ 
for small $x\ll x^*$ and by $t_f^{(s)}\simeq 2\gamma(1-x)/(xm_s)$ for 
$x\gg x^*$. 

In the parametric analysis, one has to distinguish between two distinct 
cases, namely $m_g^3>\hat{q}_g$ and $m_g^3<\hat{q}_g$. 
In perturbative QCD (pQCD), $m_g^3>\hat{q}_g$ because the in-medium gluon 
mass $m_g\sim gT$~\cite{Kampfer00,Djordjevic03,leBellac}, while 
$\hat{q}_g\sim g^4T^3$~\cite{ArnoldXiao} for small QCD running coupling 
$g\ll 1$ at large temperatures $T$. We start our discussion with this case. 
Nonetheless, we do not want to restrict the analysis of the parameter space 
to the case $m_g^3>\hat{q}_g$ only, because the opposite case could be 
satisfied in the non-perturbative (strong coupling) regime. 

For $m_g^3>\hat{q}_g$, the self-consistency condition Eq.~(\ref{eq:SCC}) implies that the minimal 
allowed $x$-value is given by $x^*/\gamma$, at which $t_f^{(s)}<t_f^{(m)}$. 
In case $\gamma<\gamma_c^{(1)}\sim m_g^3/\hat{q}_g$, 
$t_f^{(s)}(x)<t_f^{(m)}(x)$ for all $x$. Then, the formation time will 
be determined by $t_f^{(s)}$ over the whole $x$-range. In contrast, 
for $\gamma>\gamma_c^{(1)}$ a region exists, in which 
$t_f^{(s)}(x)>t_f^{(m)}(x)$. This region is bound by 
$x_1\sim x^*m_g^3/(\gamma\hat{q}_g)$ and $x_2\sim C/(1+C)$ with 
$C=(\gamma\hat{q}_g/m_s^3)^{1/3}$, at which $t_f^{(s)}=t_f^{(m)}$. This 
situation is illustrated in Fig.~\ref{Fig:4new}. The formation times at the 
intersection points read 
\begin{equation}
t_f(x_1)=t_f^{\rm onset}\simeq \frac{m_g^2}{\hat{q}_g} \,,
\end{equation} 
which is independent of $\gamma$ and represents the minimal amount of time necessary for
gluon formation in a multiple scattering process, and $t_f(x_2)\simeq (\gamma^2/\hat{q}_g)^{1/3}$, 
which increases with $\gamma^{2/3}$. 

In the case $m_g^3<\hat{q}_g$, in contrast, the picture changes. Here, 
the self-consistency condition Eq.~(\ref{eq:SCC}), supplemented by $\langle k_\perp^2 \rangle =\hat{q}_g\,t_f^{(m)}$, implies that the minimal allowed $x$-value 
is rather given by $x_{1,S}\sim\hat{q}_g^{1/3}/(\gamma m_s)$. This implies that gluons cannot be formed at rest 
with in-medium mass $m_g$ due to their reinteractions within the medium. At $x_{1,S}$ one finds 
$t_f^{(s)}>t_f^{(m)}$, with 
\begin{equation}
t_f^{(m)}(x_{1,S})=t_{f,S}^{\rm onset} \simeq\hat{q}_g^{-1/3} \,.
\end{equation} 
For any possible $\gamma$, 
$t_f^{(m)}$ is smaller than $t_f^{(s)}$ in the entire $x$-region between $x_{1,S}$ and $x_2$, so that the formation 
time is determined by $t_f^{(m)}$ in this region. In contrast, for $x>x_2$ one finds 
$t_f^{(s)}(x)<t_f^{(m)}(x)$.

\begin{table*}
\vspace{5mm}
\centering
 \begin{tabular}[t]{|l |l|l|}
  \hline
  \rule{0pt}{2ex}
  $\gamma$-scales & noticeable points & conditions \\
  \hline
  \rule{0pt}{3ex}
  $\gamma_c^{(1)}\sim m_g^3/\hat{q}_g$ & $x^*=m_g/m_s$ &  \\
  $\gamma_d^{(1)}\sim \sqrt{\hat{q}_g/\Gamma^3}$ & $x^*_S=\hat{q}^{1/3}_g/m_s$ &   \\ 
  $\gamma_d^{(2)}\sim m_g/\Gamma$ & $x_1\sim x^*(\gamma_c^{(1)}/\gamma)\sim m_g^4/(\gamma m_s\hat{q}_g)$ &  $t_f^{(s)}(x_1)=t_f^{(m)}(x_1),\,\, x\ll x^*$ \\
  & $x_{1,S}\sim x^*_S/\gamma$  & 
  \\
  & $x_2\sim C/(1+C)\sim(\gamma\hat{q}_g)^{1/3}/(m_s+(\gamma\hat{q}_g)^{1/3})$  & $t_f^{(s)}(x_2)=t_f^{(m)}(x_2),\,\, x\gg x^*$ \\
  & $x_3\sim\hat{q}_g/(\Gamma^2\gamma m_s)$  & $t_d=t_f^{(m)}(x_3)$ \\
  & $x_4\sim\gamma\Gamma/(m_s+\gamma\Gamma)$ & $t_d=t_f^{(s)}(x_4),\,\, x\gg x^*$  \\
  & $x_5\sim x^*\gamma_d^{(2)}/\gamma\sim m_g^2/(\Gamma\gamma m_s)$ & $t_d=t_f^{(s)}(x_5),\,\, x\ll x^*$
  \\
  \hline
 \end{tabular}
 \caption[]{\label{Tab:2} Summary of the definitions of the variables appearing 
 in the text and depicted in Figs.~\ref{Fig:4new},~\ref{Fig:5new},~\ref{Fig:5+1new},~\ref{Fig:6new} 
 and~\ref{Fig:9} as determined from these definitions, where $C=x^*(\gamma/\gamma_c^{(1)})^{1/3}$.}
\end{table*}
In the BDMPS-approach~\cite{Baier97}, incoherent scatterings determine the 
gluon radiation spectrum in hot QCD matter for 
$x<x_{LPM}$, where $x_{LPM}\sim\lambda_g\mu^2/(\gamma m_s)$. Here, $\lambda_g\sim(g^2T)^{-1}$ 
is the gluon mean free path between successive elastic scatterings in the 
medium, while $\mu\sim gT$ is the typical momentum transfer to the gluon in 
a single scattering. For $x>x_{LPM}$, instead, the spectrum is 
suppressed $\propto x^{-1/2}$ due to the LPM effect analogon. In our 
approach, we find in the case $m_g^3>\hat{q}_g$ that 
$x_1\sim x_{LPM}$, when we insert for the quantities entering 
$x_1$ their parametric dependencies as known from pQCD. Moreover, we 
find parametrically that $t_f(x_1)\simeq m_g^2/\hat{q}_g\sim\lambda_g$. This 
analogy between BDMPS and the case $m_g^3>\hat{q}_g$ is, however, only 
possible because $m_g$ is proportional to $\mu$ in pQCD. Naively, one would rather 
expect that the above discussed case $m_g^3<\hat{q}_g$ resembles the BDMPS-results, which 
are derived for $m_g=0$. 
Nonetheless, one cannot associate this case with BDMPS, because in our approach at vanishing $m_g$ coherence effects dominate for small $x$-values, even though the corresponding formation length is smaller than $\lambda_g$. This unphysical situation can be cured by adding the 
scale $\mu^2$ - understood as a minimal value for $\langle p_B^2\rangle$ in 
Eq.~(\ref{equ:formtimeQCD}) - 
in the denominator of $t_f^{(s)}$ in Eq.~(\ref{equ:tfsQCD}). 
Then, one finds $t_f^{(s)}<t_f^{(m)}$ for $x<\mu^4/(\hat{q}_g\gamma m_s)$, where $\mu^4/(\hat{q}_g\gamma m_s)\sim x_{LPM}$ in pQCD. 
This analysis shows that the case $m_g^3<\hat{q}_g$ does not have to be understood 
as $m_g^3$ alone being small compared to $\hat{q}_g$ but rather as the 
case of a $\hat{q}_g$ that is large compared to both $m_g^3$ and $\mu^3$. 

\subsection{Competition between gluon bremsstrahlung formation and damping \label{sec:4_2}}

We now proceed by including damping effects into our
considerations. We recall that our main symbols are defined in Tab.~\ref{Tab:2}. 
We assume that the damping mechanisms 
impose a competing time scale $t_d\simeq 1/\Gamma$ for the formation of gluon 
bremsstrahlung similar to Sec.~\ref{sec:3_2}. Then, they become influential if $1/\Gamma\lesssim t_f$. 
In Fig.~\ref{Fig:5new}, we illustrate where in $\gamma$-$x$-space damping effects are important in case $m_g^3>\hat{q}_g$. 
In Fig.~\ref{Fig:5+1new}, a larger value of $\hat{q}$ was chosen, such that $m_g^3<\hat{q}_g$. For negligible 
$\Gamma$, coherence effects dominate the radiation formation in the region between $x_1$ and $x_2$ in case $m_g^3>\hat{q}_g$ 
(cf.~Fig.~\ref{Fig:5new}), and between $x_{1,S}$ and $x_2$ in case $m_g^3<\hat{q}_g$ (cf.~Fig.~\ref{Fig:5+1new}). 
These regions increase with increasing $\gamma$ (or increasing energy $E$ for fixed $m_s$). For a non-negligible $\Gamma$, however, the situation changes. 

Considering first the case $m_g^3>\hat{q}_g$, one finds the following picture: 
For fixed damping rates $\Gamma<\hat{q}_g/m_g^2$ (case sketched in panel (a) of Fig.~\ref{Fig:5new}), 
which implies that $t_d>t_f(x_1)=t_f^{\rm onset}$, damping mechanisms become only important for 
$\gamma>\gamma_d^{(1)}\sim\sqrt{\hat{q}_g/\Gamma^3}$, i.e.~when 
$t_f^{\rm onset}<t_d<t_f(x_2)$. The scale $\gamma_d^{(1)}$
is by definition larger than $\gamma_c^{(1)}$ but decreases with increasing 
$\Gamma$. For $\gamma>\gamma_d^{(1)}$, damping effects are of significance 
in a region between $x_3$ and $x_4$ (region (I), see also in Fig.~\ref{Fig:4new} the long-dashed 
horizontal line in the middle), where $x_3\sim\hat{q}_g/(\Gamma^2\gamma m_s)$ and 
$x_4\sim\gamma\Gamma/(m_s+\gamma\Gamma)$. These boundary points are determined 
from the conditions $t_d=t_f^{(m)}(x_3)$, and from 
$t_d=t_f^{(s)}(x_4)$ for $x_4\gg x^*$, respectively. 

The lower bound $x_3$ decreases with $\gamma^{-1}$ (but slower than $x_1$) 
and will approach $x_1$ only if $\Gamma\to\hat{q}_g/m_g^2$, for which also 
$\gamma_d^{(1)}\to\gamma_c^{(1)}$. The upper bound $x_4$, instead, increases 
with $\gamma$ (but faster than $x_2$). Thus, the $x$-region in which damping 
mechanisms are of importance increases with increasing $\gamma$ and/or increasing 
$\Gamma$. 

In contrast, for larger $\Gamma>\hat{q}_g/m_g^2$ (case sketched in panel (b) of Fig.~\ref{Fig:5new}) one finds 
$t_d<t_f^{\rm onset}$ (see also in Fig.~\ref{Fig:4new} the lower long-dashed horizontal line). 
Then, damping mechanisms become important already for 
$\gamma>\gamma_d^{(2)}\sim m_g/\Gamma$, where $\gamma_d^{(2)}<\gamma_c^{(1)}$, 
in a region between $x_5\sim m_g^2/(\Gamma\gamma m_s)$ and $x_4$ (region (II)). 
The lower bound $x_5$ is determined from the condition $t_d=t_f^{(s)}(x_5)$ for $x_5\ll x^*$. 
We note that $x_5$ decreases faster than $x_1$ with increasing $\gamma$. Thus, we find for the case $\Gamma>\hat{q}_g/m_g^2$ 
that damping mechanisms influence the formation of gluon bremsstrahlung in a large region of $\gamma$-$x$-space, 
implying a negligible role of coherence effects on the radiative energy loss spectrum. 

Considering now the case $m_g^3<\hat{q}_g$, i.e.~when 
$t_f^{(m)}<t_f^{(s)}$ at $x_{1,S}$, 
we find the situation depicted in Fig.~\ref{Fig:5+1new}: 
For fixed damping rates $\Gamma<\hat{q}_g^{1/3}$, damping effects are of 
importance for $\gamma>\gamma_d^{(1)}$ in a region between $x_3$ and $x_4$ (region (I)) 
as in the above case $m_g^3>\hat{q}_g$ (cf.~panel (a) of Fig.~\ref{Fig:5new}). We conclude that, 
also in the $m_g^3<\hat{q}_g$ case, damping effects become more important for the formation of gluon 
bremsstrahlung with increasing $\gamma$ and/or $\Gamma$. As $\Gamma\to \hat{q}_g^{1/3}$, one would find 
that $x_3<x_{1,S}$ for any $\gamma$ and, thus, damping effects would be of significance in almost 
the entire $\gamma$-$x$-space. However, here we refrain from considering 
damping rates $\Gamma$ as large as $\Gamma\to \hat{q}_g^{1/3}$ or even larger. This is because 
for $\Gamma\to \hat{q}_g^{1/3}$ one finds $\Gamma>m_g$ such that the 
underlying assumption, that the inclusion of damping phenomena does not 
qualitatively influence the above $\Gamma$-independent discussion of $t_f$, would become invalid. 
In any case, we expect the effect of damping to be more pronounced with 
increasing $\Gamma$ in this regime, too. 

\subsection{Impact on the spectra \label{sec:4_4}}

\begin{figure*}[t]
\centering
\vspace{5mm}
\includegraphics[width=0.45\textwidth]{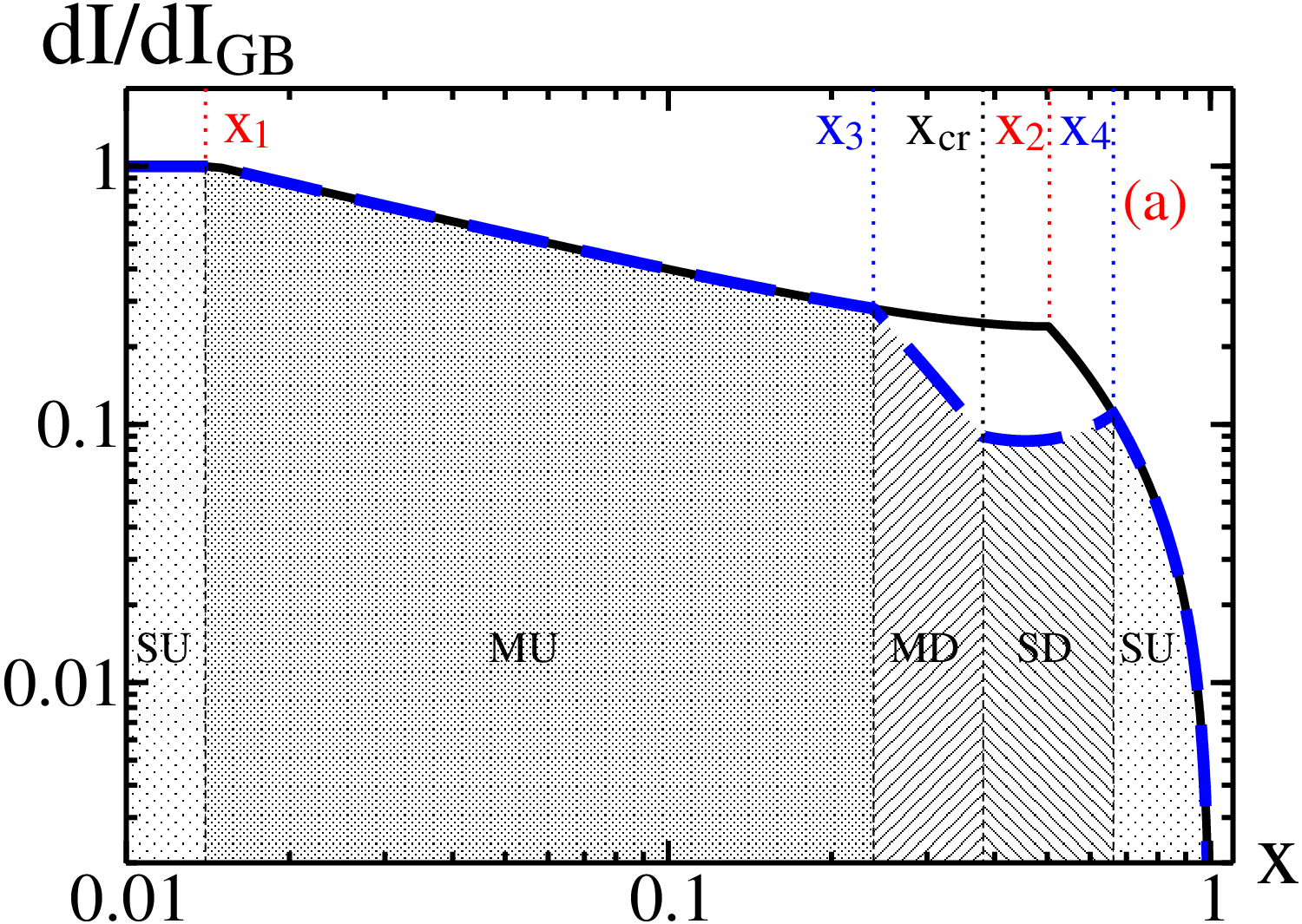} 
\hspace{3mm}
\includegraphics[width=0.45\textwidth]{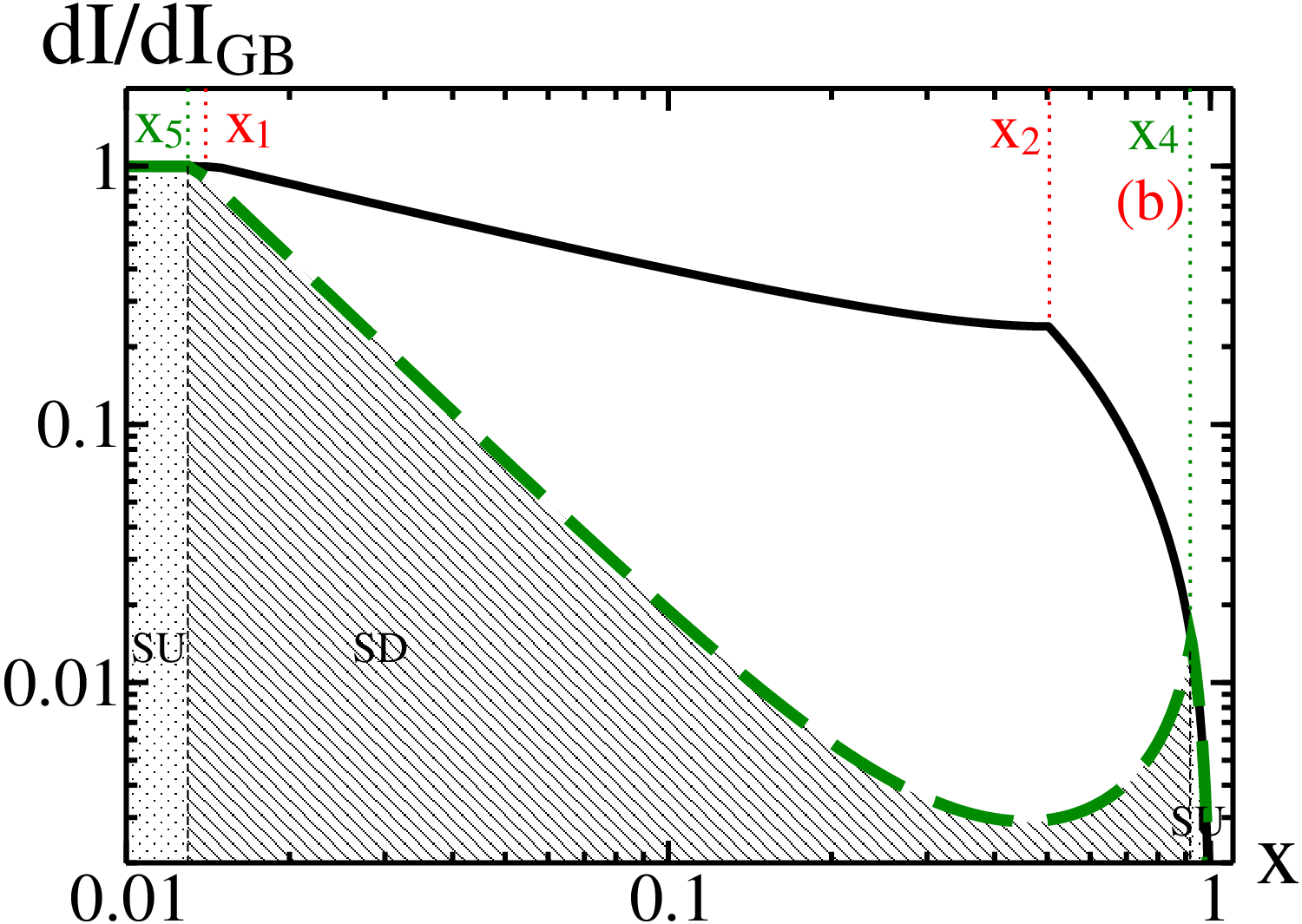} 
\caption[]{\label{Fig:6new} (Color online) (a): Visualization of the influence of gluon bremsstrahlung damping on the radiation 
spectrum relative to the GB-result for soft gluon radiation off massive partons from incoherent scatterings, 
cf.~\cite{Gossiaux10}. The parameters are chosen as $E=45$~GeV, $m_s=1.5$~GeV, $m_g=0.6$~GeV and 
$\hat{q}_g=0.2~{\rm GeV}^2/{\rm fm}$, i.e. the case $\gamma>\gamma_c^{(1)} = m_g^3/\hat{q}_g>1$ is 
considered (corresponding to the physical situation depicted in Fig.~\ref{Fig:4new}). The 
damping effect is quantified by making use of the scaling laws described in the text. The solid curve depicts 
$dI/dI_{GB}$ for $\Gamma\to 0$~GeV, while the dashed curve shows $dI/dI_{GB}$ for $\Gamma=0.055$~GeV, for which 
$\Gamma<\hat{q}_g/m_g^2$ (corresponding to the situation illustrated by the middle long-dashed horizontal line in 
Fig.~\ref{Fig:4new}). The different shaded regions are described in the text. (b): As in panel 
(a), but for $\Gamma=0.3$~GeV, for which $\Gamma>\hat{q}_g/m_g^2$ (corresponding to the situation depicted by the lower 
long-dashed horizontal line in Fig.~\ref{Fig:4new}). This highlights the increasing influence of damping effects 
on the radiation spectrum with increasing $\Gamma$.} 
\end{figure*} 
The above formation and damping time analysis allows us to discuss qualitatively 
the influence of different in-medium effects such as damping on the radiation spectrum in QCD. 
Given the generic structure of the spectrum discussed in Sec.~\ref{sec:2}, we make use of the scaling 
laws determined there. However, here we want to analyze the behaviour of the radiative power spectrum  $dI/dx$ relative to the 
soft, i.e.~$x\ll x^*$, Gunion-Bertsch (GB) power spectrum limit 
of gluon radiation off massive partons $dI_{GB}/dx$ from incoherent 
scatterings. The latter was determined within scalar QCD in~\cite{Gossiaux10} as an extension of the result by Gunion and 
Bertsch~\cite{Gunion82} for massless partons. With this normalization, the analogon of the scaling law in 
Eq.~(\ref{eq:small_damping}) reads as 
\begin{equation}
\label{equ:scalingQCD1}
 \frac{dI}{dI_{GB}} \simeq \kappa\frac{t_f}{t_{GB}} 
\end{equation}
with $t_f\simeq\min\{t_f^{(s)}, t_f^{(m)}\}$. The scale $t_{GB}\simeq 2x\gamma m_s/m_g^2$ is the 
formation time $t_f^{(s)}$ of soft gluon radiation in a single scattering process. 

Considering first the case $m_g^3>\hat{q}_g$, the spectra ratio is $dI/dI_{GB}=1$ for small $x$ 
as seen in the small-$x$ regions (SU) in Fig.~\ref{Fig:6new}, because $t_f$ is given by $t_f^{(s)}$ 
for such $x$ (cf.~also Fig.~\ref{Fig:4new}). With increasing $x$, the ratio 
$dI/dI_{GB}$ is reduced, where the reduction is proportional to a 
specific power of the gluon fractional energy in line with the appropriate scaling law. 

For negligible damping rates $\Gamma$, as illustrated by solid curves in Fig.~\ref{Fig:6new} panels (a) and (b), 
$dI/dI_{GB}\propto (x(1-x))^{-1/2}$ if $t_f^{(m)}<t_f^{(s)}$, which holds for $x_1<x<x_2$. This includes, in particular, the known 
BDMPS-Z suppression of the power 
spectrum~\cite{Baier95,Baier97,Zakharov} $dI/dI_{GB}\propto x^{-1/2}$ for $x\ll 1$. If, instead $x>x_2$, 
i.e.~$t_f^{(s)}<t_f^{(m)}$, then $dI/dI_{GB}\propto (1-x)/x^2$ is found. 
This behaviour is a consequence of the finite parton mass $m_s$ 
in Eq.~(\ref{equ:tfsQCD}) and, thus, a feature specific to heavy quarks, cf.~\cite{Peigne}. 

For non-negligible damping rates $\Gamma$, one finds at most four different physical regions and sizeable parts of the 
spectrum can become additionally reduced. For $\Gamma<\hat{q}_g/m_g^2$ (see dashed curve in Fig.~\ref{Fig:6new} panel (a)), damping 
effects influence the regions at intermediate $x$, that were dominated by multiple undamped (MU) and single undamped (SU) scatterings 
in the absence of damping. For $x_{cr}<x<x_2$, damping effects are so strong that effectively the formation of radiation is hampered 
already after a single scattering process, even though $t_f^{(m)}<t_f^{(s)}$ in this region. Here, $x_{cr}$ is determined from $t_d=t_f^{(m)\,2}/t_f^{(s)}$, and follows, assuming $x_{cr}>x^*$, as $x_{cr}\sim 1/(1+m_s\sqrt{\Gamma/\hat{q}_g})$.
Passing from small $x$-values towards 
$x=1$, thus, the spectrum is dominated first by single undamped scatterings, then by multiple undamped scattering processes with the 
known BDMPS-Z suppression, followed by multiple damped (MD) processes, where the spectrum suppression is $\propto (1-x)/x^2$ in line 
with 
\begin{equation}
 \label{equ:32}
 \frac{dI}{dI_{GB}} \simeq \frac{t_d^3}{t_f^{(m)\,2}\, t_{GB}} \,. 
\end{equation}
For $x_{cr}<x<x_2$, single damped (SD) scatterings dominate the spectrum with 
a modification 
factor $\propto 1/(1-x)+m_g^2/(x^2m_s^2)$ in line with 
\begin{equation}
 \label{equ:33}
 \frac{dI}{dI_{GB}} \simeq \frac{t_d^2}{t_f^{(s)}\,t_{GB}} \,, 
\end{equation}
and for $x>x_2$ a 
region (SU) follows with $dI/dI_{GB}\propto (1-x)/x^2$. 
The spectrum suppression in the damped regions (MD) and (SD) is evidently stronger 
than the suppression $\propto x^{-1/2}$ in the BDMPS-Z regime. 

With increasing $\Gamma$, the regions dominated by damping mechanisms increase. For $\Gamma>\hat{q}_g/m_g^2$, the 
physical situation changes significantly (cf.~panel (b) in Fig.~\ref{Fig:6new}). In a large $x$-region between $x_5$ and $x_4$ 
(cf.~also Fig.~\ref{Fig:5new} panel (b)) single damped scatterings determine the power spectrum according to Eq.~(\ref{equ:33}), and there is no room left for coherence effects. It is noteworthy, however, that for larger $x$ the term $\propto 1/(1-x)$ dominates in Eq.~(\ref{equ:33}) such that the effect of a power spectrum suppression due to damping becomes reduced. 

\begin{figure}[t]
\centering
\vspace{5mm}
\includegraphics[width=0.45\textwidth]{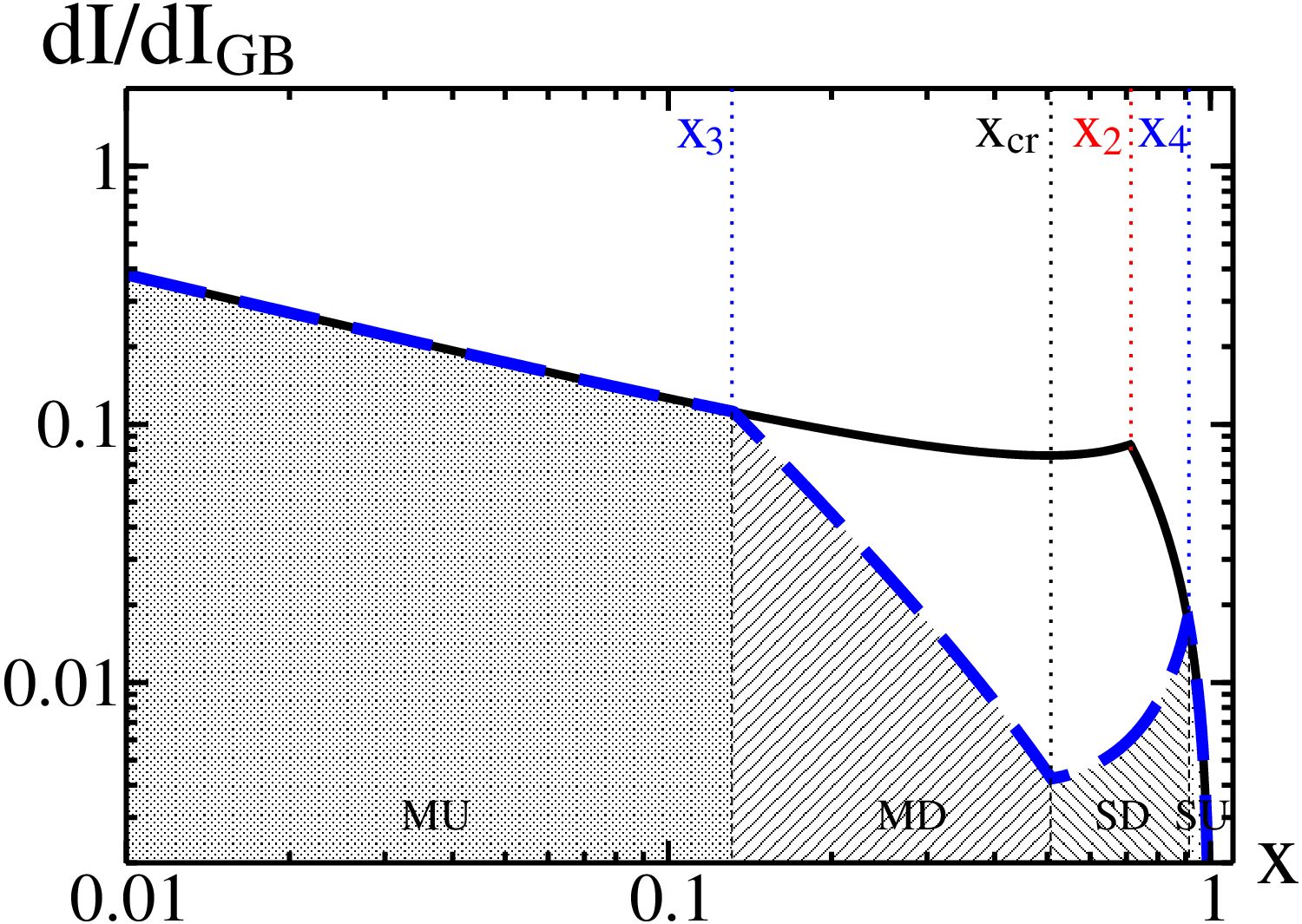} 
\caption[]{\label{Fig:9} (Color online): As in Fig.~\ref{Fig:6new}, but for $\hat{q}_g=2$~GeV$^2/$fm, i.e. for 
the case $m_g^3<\hat{q}_g$. The solid curve shows the result for negligible $\Gamma$, while 
the dashed curve exhibits $dI/dI_{GB}$ for $\Gamma=0.25$~GeV, for which $\Gamma<\hat{q}_g^{1/3}$. The different 
shaded regions are discussed in the text.} 
\end{figure} 
The opposite case $m_g^3<\hat{q}_g$ is shown in Fig.~\ref{Fig:9}. For small $x$, multiple undamped scattering processes 
dominate the power
spectrum and $dI/dI_{GB}<1$ already in this region. For negligible $\Gamma$, the physical situation changes only for $x>x_2$, 
where single undamped scatterings determine the spectrum. For non-negligible $\Gamma<\hat{q}_g^{1/3}$, damping effects become influential 
at intermediate $x$, where for $x<x_{cr}$ a region of multiple damped scatterings and for $x>x_{cr}$ a region of single damped 
scatterings emerges. Again, the spectrum suppression in the regions (MD) and (SD) is stronger than in the regime (MU), in which coherence 
effects dominate the radiation spectrum. 

\subsection{Discussion of possible damping mechanisms \label{sec:4_3}}

Damping phenomena specific to hot QCD matter become important for the radiative energy 
loss of an energetic parton, when they start to influence the formation process 
of the bremsstrahlung gluons. The gluon damping rate is related to the imaginary 
part of the poles in the corresponding in-medium propagator~\cite{Weldon83}. 
In perturbative QCD approaches, it has been calculated both for collective gluon 
modes in the plasma~\cite{Braaten90,Thoma} as well as for hard 
gluons~\cite{Pisarski93}. For collective gluons with momenta and energies of 
$\mathcal{O}(gT)$ both elastic and inelastic processes contribute equally to 
a damping rate $\Gamma\sim g^2T$. In case of a hard gluon with either momentum 
or energy of $\mathcal{O}(T)$ or larger only elastic processes have been 
considered in the evaluation of the gluon damping rate in~\cite{Pisarski93}, yielding 
$\Gamma\sim g^2T\ln(1/g)$ for $g\ll 1$. However, elastic rescatterings 
of a bremsstrahlung gluon during its formation are already taken into account 
by $\hat{q}_g$ in our considerations. The nature of the damping mechanisms we have in mind here is 
different: As possible inelastic processes leading to a damping of gluon 
radiation, one might consider either quark--anti-quark pair production or 
secondary bremsstrahlung creation from a gluon during its formation process. 

For estimating the parametric dependence of the latter process, one might 
view the emission of the secondary bremsstrahlung gluon in the 
dense medium as a BDMPS-mechanism~\cite{Baier97}. In doing so, a preformed 
gluon with fractional energy $x>x_{LPM}$, which is, however, small compared to $1$, 
emits a secondary bremsstrahlung gluon of fractional energy $x'$ in an 
inelastic scattering process. In order to be of influence on the formation 
of the preformed gluon, the formation time of the secondary gluon must be small 
compared to the formation time of its emitter, which can be realized if 
$x'\ll x$, cf.~Fig.~\ref{Fig:4new}. The production rate for the 
secondary gluon as a measure for the damping rate of the preformed emitter gluon is 
obtained by integrating the weighted BDMPS-radiation spectrum~\cite{Baier97}. 
Neglecting the part in the integral, which is suppressed with 
$x^{-3/2}$, one finds a gluon damping rate $\Gamma\sim g^4T$ up to 
corrections of order $\ln(1/g)$. We note that a similar parametric 
dependence was found in~\cite{Biro93} for the gluon production rate, which is of 
interest for the chemical equilibration of the hot QCD plasma. For such a 
damping rate, the case $\Gamma<\hat{q}_g/m_g^2$ would be realized in pQCD, 
leading to the situation depicted in panel (a) of Fig.~\ref{Fig:5new}. 

\section{Conclusion \label{sec:5}}

Damping phenomena in a dense, absorptive plasma can influence substantially 
the radiative energy loss of an energetic charge traversing this medium. 
They manifest themselves in a non-trivial reduction of the associated radiation spectrum 
off that charge as compared 
to the spectrum from incoherent scatterings in non-absorptive matter. The 
effect is more pronounced for large energies $E$ 
(or equivalently large Lorentz-factors $\gamma$) of the charge and/or large damping 
rates $\Gamma$ of the radiated quanta in the medium. This behaviour can be 
understood semi-quantitatively by making use of the concept of a 
formation time for radiation: Damping mechanisms reduce the radiative energy loss 
spectrum if the typical time scale for the damping of radiation quanta with fractional 
energy $x=\omega/E$ is small compared to their formation time such that 
these effects influence already the creation of the radiation. 

In this article, we analyzed systematically the interplay between these 
competing time scales for an absorptive QED and QCD medium. We started with the 
case of a polarizable and absorptive, infinite electro-magnetic plasma, for 
which analytical results for the radiative energy loss spectrum per unit length 
have been derived in~\cite{Bluhm11,Bluhm12}. Then, we extended the phenomenological discussion to QCD 
matter. This is the first time that the consequences of damping are studied in QCD 
as in perturbative QCD approaches damping phenomena were so far considered as negligible, higher-order effects. 

In both cases, i.e.~QED and QCD, we identified parametrically the regions in 
$\gamma$-$x$-space, in which either coherence or damping effects 
significantly influence the associated radiation spectrum, 
cf.~Figs.~\ref{Fig:2newPLUS},~\ref{Fig:5new}~and~\ref{Fig:5+1new}. 
We showed that, generically, damping effects become important 
in an intermediate-$x$ regime, which grows with increasing $\gamma$ and/or $\Gamma$. 
Any suppression of the spectrum in this regime has to 
be attributed to damping phenomena rather than coherence effects. 
We showed that the radiation 
spectrum is stronger suppressed through damping effects than it is through coherence effects. 
Restricting our analysis for QCD matter to the case of heavy quarks, we visualized 
this feature in Figs.~\ref{Fig:6new} and~\ref{Fig:9} 
by using typical values for the entering parameters. 
The utilized concept should, however, be generalizable to the 
study of light partons as well. 

As damping effects become pronounced for large $\gamma$, the study of hadronic 
correlations at high transverse momenta $p_T$ might open the avenue for experimentally 
measuring the absorptive properties of hot QCD matter. Moreover, the effect could 
be a key ingredient for the understanding of the observed heavy 
meson spectra: If $t_d$ becomes the dominant scale, the radiation spectra 
turn out to be mass-independent, which would constitute a 
step towards solving the non-photonic single-electron puzzle~\cite{Djordjevic:2006kw}. 
Corresponding phenomenological investigations are presented in~\cite{Nahrgang}, where the quark-mass 
independence of the heavy-flavor meson quenching at large $p_T$ is quantified and a comparison with the available 
experimental data from the ALICE collaboration on the $D$ meson quenching suggests a rather strong damping rate 
$\Gamma/T\simeq 0.75$. Also, 
we expect our results to be of relevance 
in heavy-quark tagged jet physics: On the one hand, either the finite in-medium gluon mass 
$m_g$ or the parameter $\hat{q}_g^{1/3}$ represent a natural lower energy cutoff for 
bremsstrahlung gluons in the soft region of the spectrum. Damping effects, 
on the other hand, hamper the formation of hard or intermediate hard gluons, 
providing effectively an upper limit for the radiative energy loss of an 
energetic projectile, similar to coherence effects but with a stronger impact on the power spectra. 

In this work, we neglected the consideration of any dependence on a finite 
path length $L$. Nevertheless, whether the energetic charge is created in the remote past or originating from the interior 
bulk of a thick medium, 
only quanta with a formation length smaller than the distance $L$ travelled by the charge 
can contribute to the medium-induced radiative energy loss, cf.~\cite{Peigne}. This implies that in the regions 
of $\gamma$-$x$-space, in which damping effects are relevant, finite path length effects play no additional role 
for the radiation spectrum off these charges. Following 
the heuristic arguments presented 
in~\cite{Baier97}, we expect that the averaged radiative energy loss of such 
partons in the hot QCD plasma is proportional to $L/\Gamma$ in case $1/\Gamma<L$. 
Only for partons stemming from the outer crust of the medium of thickness $1/\Gamma$ finite path length effects 
are important and the averaged radiative energy loss is proportional to $L^2$. 
We considered, moreover, $x$-independent 
damping rates in our analysis. In general, however, $\Gamma$ should depend 
on the energy of the radiation quanta. We leave such studies for future investigations. 

\section*{Acknowledgements}
We acknowledge valuable discussions with Yu.~L.~Dokshitzer, E.~Iancu, 
B.~K\"ampfer, S.~Peign\'{e} and M.~H.~Thoma. We also thank Yu.~A.~Markov 
for drawing our attention to Ref.~\cite{Galitsky}. 
The work is supported by the European Network I3-HP2 Toric, 
the ANR research program ``Hadrons@LHC'' (grant ANR-08-BLAN-0093-02) 
and the ``Pays de la Loire'' research project TOGETHER. 

\end{document}